\documentclass[AMA,STIX1COL,doublespace]{WileyNJD-v2}

\articletype{Article Type}%

\received{}
\revised{}
\accepted{}
        
\begin{document}

\title{Propensity scores using missingness pattern information: a practical guide}

\author[1]{Helen A.\ Blake*}
\author[1,2]{Cl\'{e}mence Leyrat}
\author[2]{Kathryn E.\ Mansfield}
\author[3]{Shaun Seaman}
\author[2]{Laurie A.\ Tomlinson}
\author[1,4]{James Carpenter}
\author[1,5]{Elizabeth J.\ Williamson}

\authormark{BLAKE \textsc{et al}}

\address[1]{\orgdiv{Department of Medical Statistics}, \orgname{London School of Hygiene \& Tropical Medicine}, \orgaddress{\state{London}, \country{UK}}}
\address[2]{\orgdiv{Department of Non-communicable Disease Epidemiology}, \orgname{London School of Hygiene \& Tropical Medicine}, \orgaddress{\state{London}, \country{UK}}}
\address[3]{\orgdiv{MRC Biostatistics Unit}, \orgname{School of Clinical Medicine}, \orgaddress{\state{Cambridge}, \country{UK}}}
\address[4]{\orgdiv{MRC Clinical Trials Unit}, \orgname{University College London}, \orgaddress{\state{London}, \country{UK}}}
\address[5]{\orgname{Health Data Research UK}, \orgaddress{\state{London}, \country{UK}}}

\corres{*Helen Blake, London School of Hygiene \& Tropical Medicine, Keppel Street, London WC1E 7HT, UK \\ \email{Helen.Blake@lshtm.ac.uk}}

\abstract[Abstract]{Electronic health records are a valuable data source for investigating health-related questions, and propensity score analysis has become an increasingly popular approach to address confounding bias in such investigations. However, because electronic health records are typically routinely recorded as part of standard clinical care, there are often missing values, particularly for potential confounders. In our motivating study -- using electronic health records to investigate the effect of renin-angiotensin system blockers on the risk of acute kidney injury -- two key confounders, ethnicity and chronic kidney disease stage, have 59\% and 53\% missing data, respectively. 
	
The missingness pattern approach (MPA), a variant of the missing indicator approach, has been proposed as a method for handling partially observed confounders in propensity score analysis. In the MPA, propensity scores are estimated separately for each missingness pattern present in the data. Although the assumptions underlying the validity of the MPA are stated in the literature, it can be difficult in practice to assess their plausibility.
	
In this paper, we explore the MPA's underlying assumptions by using causal diagrams to assess their plausibility in a range of simple scenarios, drawing general conclusions about situations in which they are likely to be violated. We present a framework providing practical guidance for assessing whether the MPA's assumptions are plausible in a particular setting and thus deciding when the MPA is appropriate. We apply our framework to our motivating study, showing that the MPA's underlying assumptions appear reasonable, and we demonstrate the application of MPA to this study. 
}

\keywords{electronic health records, missing confounder data, missing indicator, missingness pattern, propensity score analysis}

\maketitle

\section{Introduction}

Observational data are an important source of information for investigating the effect of treatments or interventions on health outcomes. In observational data, confounding is often an issue, as characteristics of treated patients can systematically differ from those of untreated patients. Propensity score methods aim to take account of confounding by achieving balance of patient characteristics across the treatment groups being compared \cite{Ros83}. However, observational studies may suffer from large amounts of missing data, which can lead to biased treatment effect estimates if the missing data are not handled appropriately \cite{Bar15}. We focus on scenarios where the outcome and treatment of interest are fully observed, but data are missing on potential confounders. This is a common occurrence, for example, in studies using electronic health record data and insurance claims data, where prescriptions and diagnoses tend to be well recorded but potential confounders, such as smoking status, may be less well recorded\cite{Her15}.\\

The 'missingness pattern' approach (MPA) is a way of handling missing confounder data that has been proposed in propensity score analysis \cite{Ros84,DAg00}. It accounts for missing data by incorporating information about which confounders are missing into the estimation of the propensity score itself \cite{Ros84,DAg00}. Despite being easy to implement, the MPA has not been widely used in practice. This might be explained by the lack of guidance about its use in the literature. In particular, while the assumptions required for the validity of the MPA have been described formally in terms of conditional independence \cite{Ros84,DAg00,Mat09}, how these mathematical statements relate to real clinical scenarios remains unclear. Our aim is therefore to investigate  the assumptions underlying the MPA in order to provide practical guidance for researchers about how to identify whether these assumptions hold in a given clinical scenario.\\

We start by introducing our motivating example which investigates the association between renin-angiotensin system drugs and risk of acute kidney injury in Section \ref{sec:illustrateAKI}. 
We review propensity score methods for complete data (Section \ref{sec:b/g}) and approaches to handle missing confounder data in propensity score analysis, with a particular focus on the MPA and its underlying assumptions and the MPA's connection to the commonly used missing indicator approach (Section \ref{sec:PSAwithMD}). 
We explore the assumptions underlying the MPA in Section \ref{sec:investigatingMPAassmptns} by considering settings where the MPA's assumptions seem plausible. We also consider how we can use causal diagrams to assess the plausibility of these assumptions in Section \ref{sec:method} and present a framework giving practical guidance for assessing the assumptions in Section \ref{sec:practical_guidance}. 
We illustrate our results on our motivating example (Section \ref{sec:results_eg}) and conclude with a discussion (Section \ref{sec:discussion}).

\section{Motivating Example}\label{sec:illustrateAKI}
We consider data from a cohort study using electronic health records to investigate the association between use of angiotensin-converting enzyme inhibitors and angiotensin receptor blockers (ACEI/ARBs) and risk of acute kidney injury (AKI) in new users of antihypertensive drugs \cite{Man16}. 

Data were obtained from the UK Clinical Practice Research Datalink linked to the Hospital Episode Statistics database for adults who were new users of antihypertensive drugs between 1997-2014. Follow-up began at the first prescription of any of the antihypertensive drugs: ACEI/ARBs, beta blockers, calcium channel blockers or diuretics. Our treatment of interest is ACEI/ARB prescription at the start of follow-up, and the outcome is AKI within 5 years. Potential confounders considered are: gender; age; ethnicity; prescription of other antihypertensive drugs at start of follow-up; and status of chronic comorbidities at start of follow-up, including chronic kidney disease (CKD) stage. A summary of the baseline characteristics of the cohort is presented in Table \ref{tab:akichar}. 
Of the 570,586 patients included in the cohort, 159,389 (27.9\%) were prescribed an ACEI/ARB. Many of the characteristics presented in Table \ref{tab:akichar} are not balanced across the treatment groups, indicating potential for confounding. Propensity score analysis is a popular method for taking account of confounding in analysis of electronic health records. However, two potential confounders have missing data: ethnicity (59.0\% missing) and baseline CKD stage (52.9\% missing). Only 121,527 (21\%) of patients have complete data for both variables. The next two sections review propensity score methods and approaches that could be used to handle these missing confounder data.

\section{Propensity score methods for complete data}\label{sec:b/g}

Before we consider how to handle missing confounder data in propensity score analysis, we will first review propensity score methods for complete data.

\subsection{Notation and assumptions}\label{sec:causalinf}
Suppose we have a group of $n$ patients, each with a row vector $X_i$ of $p$ confounders: $X_i=(X_{i1},...,X_{ip})^{\top}$, where $X_{ij}$ is the value of confounder $j$ for patient $i$ ($i=1,\ldots,n$ and $j=1,\ldots,p$). Throughout the paper, we will assume that in the full data (i.e. with no missing confounder data) the $X_i$ are sufficient to control for confounding \cite{Van13}. In this paper, we restrict our attention to a binary treatment (or exposure or intervention) and a binary outcome. Each patient $i$ receives either treatment $Z_i=1$ or control $Z_i=0$. To define the causal effect of treatment as opposed to control, we use the potential outcomes framework \cite{Guo10}. Each patient has two potential outcomes: $Y_i(1)$ denotes the outcome that would have been observed for patient $i$ if they had received treatment, and $Y_i(0)$ denotes the outcome value that would have been observed if patient $i$ had received control. $Y_i$ denotes the outcome value that was actually observed. Henceforth, we omit the $i$ and $j$ subscripts where unambiguous. We express the causal effect of the treatment on the binary outcome as the risk difference to avoid the issue of the non-collapsibility of the odds ratio \cite{Gre99}; our estimand is the average treatment effect (ATE): $E[Y(1)-Y(0)]$ \cite{Wil14introps,Imb04}.\\

To estimate the treatment effect we make four standard assumptions: consistency, no interference, strongly ignorable treatment assignment (SITA), and positivity. Consistency states that, for a patient who receives a particular treatment level $z$, their observed outcome $Y$ is the corresponding potential outcome $Y(z)$, irrespective of the way in which they were assigned to that treatment level \cite{Her16}. Under the assumption of no interference, the treatment received by one patient does not affect the potential outcomes of another patient \cite{Ros83,Hof05,Hal95}. SITA implies that there are no unmeasured confounders and can be expressed as \cite{Ros83,Wil12}: 
\begin{equation}\label{eq:SITA}
\mathrm{SITA: } \hspace{0.2in} Z \perp \big(Y(1),Y(0)\big)  | X 
\end{equation} 
where $\perp$ denotes independence. Finally, positivity states that, given their individual characteristics, all patients have a non-zero probability of receiving either treatment or control \cite{Wil12,Lit00}. Throughout this paper,  we assume these four assumptions hold for the complete data.

\subsection{Propensity scores}
The propensity score $e(x)$ is the probability of receiving treatment, conditional on observed confounders $X$ \cite{Ros83}: 
\begin{equation*}
e_i(x_{i})=P(Z_i=1|X_{i}=x_{i}) \hspace{0.01in},
\end{equation*}
for patient $i$ $(i=1,\ldots,n)$ with a vector of confounder values $X_{i}=x_{i}$.
Under the four assumptions described above, Rosenbaum and Rubin \cite{Ros83} showed that at each value of the propensity score the confounders $X$ are balanced across treatment groups (i.e.\ have similar distributions in the two groups).

Typically, propensity scores are unknown and must be estimated from the data \cite{Aus11}. Often, the individual estimated propensity scores $\hat{e}_i$ are the predictions obtained from a logistic regression of treatment allocation on potential confounders \cite{Aus11}. The estimated propensity score can be used to match treated and control patients to create comparable groups, allowing a direct comparison of the mean outcomes between groups. Another approach uses the propensity score to create strata of relatively homogeneous patients and estimates the ATE by pooling the treatment effect estimated within each stratum. Alternatively, the estimated propensity score can be incorporated into a regression model for outcome on treatment \cite{Ros83}. A slightly different approach, inverse probability of treatment weighting (IPTW), uses the estimated propensity scores as weights to construct `pseudo-populations' \cite{Her06} in which the distribution of observed confounders are balanced across treatment groups \cite{Aus11}. We will focus on the IPTW method, which results in the following estimator for the ATE: 
\begin{equation}\label{eq:iptw}
\widehat{\text{ATE}} = \left(\dfrac{\sum_{i=1}^{n}\frac{Y_iZ_i}{\hat{e}_i}}{\sum_{i=1}^{n}\frac{Z_i}{\hat{e}_i}}\right) - \left(\dfrac{\sum_{i=1}^{n}\frac{Y_i(1-Z_i)}{(1-\hat{e}_i)}}{\sum_{i=1}^{n}\frac{(1-Z_i)}{(1-\hat{e}_i)}}\right) \hspace{0.01in}.
\end{equation}

\section{Propensity score methods with missing confounder data}\label{sec:PSAwithMD}

In practice, observational studies can suffer from large amounts of missing data, potentially leading to both loss of efficiency and biased estimates \cite{Car13}. The extent of any bias will depend on the extent to which the probability of missing confounder data is related to outcome and exposure \cite{Car13}. The most common classification of missingness mechanisms, i.e.\ the way in which data becomes missing, is Rubin's taxonomy, in which data are missing completely at random (MCAR), missing at random (MAR) or missing not at random (MNAR) \cite{Car13,Sea13}.
Under a MCAR mechanism, the probability of being missing does not depend on the observed or missing data. Missing data are MAR when the probability of being missing depends on observed data values but, given these, does not depend on missing values. If the probability of being missing depends on the unobserved values of data then data are MNAR. 

\subsection{Simple approaches}
The simplest way of handling missing confounder data in propensity score analysis is to perform a complete records analysis (also referred to a complete case analysis), which restricts the analyses to patients with full data on all variables \cite{Bar15}. This approach provides unbiased estimates of the average treatment effect as long as missingness does not depend on both the treatment and the outcome \cite{Bar15}. 

The missing indicator method is another simple approach. For partially observed categorical confounders, it simply involves adding a `missing' category before including the confounder in the propensity score model. For continuous confounders, missing values are set to a particular value, say 0, and both the confounder and a missingness indicator (a variable indicating whether that variable is  observed or not) are included in the propensity score model. While, if the substantive analysis is a standard outcome regression model this method is biased in a number of scenarios \cite{Gro12,Jon96}, whether this is the case in the propensity score context has been questioned \cite{DAg01}. We discuss this missing indicator approach further in Section \ref{sec:miss_ind}.

\subsection{Multiple Imputation}
An alternative, increasingly popular, approach is Multiple Imputation (MI). This involves imputing (i.e.\ filling in) missing covariates with plausible values several times, by drawing from the predictive distribution of the missing covariates given observed data, to create a number of `complete' imputed datasets. The full analysis (here, estimation of the propensity score followed by estimation of the treatment effect) is  performed separately in each imputed dataset. The results are then combined using Rubin's rules to obtain an overall estimate of the treatment effect and standard error \cite{Rub76,Car13}. Guidelines regarding optimal use of MI in conjunction with propensity score analysis have been proposed \cite{Ley17}. Standard implementations of MI require data to be missing at random\cite{Car13,Ste09}. \\

\subsection{The Missingness Pattern Approach (MPA)}

The Missingness Pattern Approach (MPA) \cite{Ros84,DAg00} accounts for missing confounder data by separating patients into subgroups according to the possible combinations of confounders being observed or missing, i.e.\ the missingness patterns, and fitting a different propensity score model to each pattern. 

Let $R_{ij}$ be a missingness indicator indicating whether the confounder $j$ $(j=1,\ldots,p)$ for patient $i$ $(i=1,\ldots,n)$ is observed ($R_{ij}=1$) or not ($R_{ij}=0)$. This allows us to partition the values $X_{ij}$ $(i=1,\ldots,n; j=1,\ldots,p)$ into two sets: the set of values that are observed, $X_{obs}$, and the set of values that are missing, $X_{mis}$:
\begin{equation}\label{eq:XobsXmis}
X=\{X_{obs}, X_{mis}\} \text{ where } X_{obs}=\{X_{ij}|R_{ij}=1\} \text{ and } X_{mis}=\{X_{ij}|R_{ij}=0\} \hspace{0.01in}.
\end{equation}
We will use $R_i = (R_{i1},...,R_{ip})$ to refer to the vector of missingness indicators for patient $i$, omitting the subscript $i$ where unambiguous.

The generalized propensity score, $e^{\ast}(x)$, is defined as the probability of receiving treatment, conditional on both the observed confounder information and the missingness pattern: $e^{\ast}(x)=P(Z=1|X_{obs},R)$. This can be estimated by using a different propensity score model for each missingness pattern, including only the confounders observed in that pattern. The estimated generalised propensity score can then be used in IPTW (or other propensity score methods) just as the standard propensity score would be used \cite{Ros84}. For example, consider a study investigating the relationship between a treatment and an outcome (both fully observed) in which there is only a single partially observed confounder $X$. Here, there are two missingness patterns: $X$ is either observed or missing. For patients with $X$ observed, the propensity score model would include $X$, whilst the propensity score model for patients with $X$ missing would include only a constant term. Each propensity score model would be fitted and used to estimate propensity scores for patients within that pattern; these are collated into a single variable (the generalised propensity score).  The ATE could then be estimated by using equation \eqref{eq:iptw} with this generalised propensity score.\\

\subsubsection{Assumptions of the Missingness Pattern Approach}
Three assumptions underlying the validity of the MPA are given by Mattei (2009) \cite{Mat09}. In this paper, we present slightly weaker versions of these assumptions, under which the MPA still gives a consistent estimator of the ATE (Appendix \ref{sec:appx:weak}). The first assumption is an extension of the SITA assumption (equation \eqref{eq:SITA}), which we call the Missingness Strongly Ignorable Treatment Assignment (mSITA) assumption due to its similarities with the SITA assumption (equation \eqref{eq:SITA}):
\begin{equation}\label{eq:mSITA}
\text{mSITA: } \hspace{0.2in} Z \perp Y(z) | X, R  \qquad \mbox{for} \qquad z=0,1.
\end{equation}
A key difference with equation \eqref{eq:SITA} is the inclusion of information about the missingness pattern, represented by $R$, in the conditioning set. We assume that SITA holds in the full data, thus this assumption states that additionally conditioning on $R$ does not introduce bias. 

Mattei presents two further assumptions \cite{Mat09}. We call these the conditionally independent treatment (CIT) assumption and the conditionally independent outcomes (CIO) assumption. As above, we present a slightly weaker version of the latter assumption. 
\begin{subequations}\label{eq:CITandCIO}
	\begin{align}
	\text{CIT: } & & Z & \perp X_{mis}|X_{obs}, R \label{eq:CIT} \hspace{1.8in} \\
	\text{CIO: } & & Y(z) & \perp X_{mis}|X_{obs}, R \hspace{0.1in} \qquad \mbox{for}\qquad z=0,1.  \label{eq:CIO} \hspace{1.2in}
	\end{align}
\end{subequations}
While these assumptions bear some superficial resemblance to the MCAR, MAR and MNAR mechanisms, unlike the latter, the former are not about the reasons why data are missing but rather about the relationships between missing confounder values and the treatment or outcome.

If mSITA holds, and either CIT or CIO holds, then the MPA provides a consistent estimate of the treatment effect. How to assess the plausibility of these assumptions --- and thus the validity of the MPA itself --- in a particular setting, however, remains unclear. In this paper, we will explore the implications of these assumptions by considering examples where they seem plausible (Section \ref{sec:investigatingMPAassmptns}), utilising causal diagrams to detect violations of these assumptions (Section \ref{sec:method}), and providing practical guidance for assessing the plausibility of these assumptions (Section \ref{sec:practical_guidance}).

We note that the assumptions underlying the MPA are different to Rubin's taxonomy of missing data \cite{Rub76,Lit02} in the sense that classifying data according to Rubin's taxonomy does not give us any information as to whether the MPA's assumptions would hold (because both $X$ and $R$ are conditioned upon). Rather, the validity of the MPA's assumptions depends on the associations between variables being different in different missingness patterns. So, as long as mSITA and either CIT or CIO holds, the MPA can be appropriate even where data is MNAR.

\subsubsection{The connection between the missingness pattern approach and the missing indicator approach}\label{sec:miss_ind}
The MPA approach is related to the missing indicator method, another commonly-used missing data approach, which involves setting missing values to a particular value and adding a missingness indicator to the propensity score model.

In a scenario with a single binary or continuous partially observed confounder, $X$, the propensity score model for the MPA  can be written as: 
\begin{equation*}
logit(P(Z=1)) = \left\{
\begin{array}{lr}
\alpha_1 + \beta_1X &\text{if } R=1 \\
\alpha_0 &\text{if } R=0. \\
\end{array}
\right.
\end{equation*}
Defining a new variable $X^*$ which takes the value $X$ if observed, and $0$ otherwise, this can be rewritten as:
\begin{equation*}
logit(P(Z=1))=  \alpha_0 + \beta_1 X^* R + (\alpha_1-\alpha_0)R, 
\end{equation*}
If $X$ is binary, this is equivalent to creating a third category for $X$ representing the missing values. If $X$ is continuous, this sets missing values to $0$ and adds an indicator variable for missing observations. This is exactly the missing indicator procedure. If $X$ were categorical, this could be extended to show that the MPA is similarly equivalent to adding a `missing' category. Since the MPA and the missing indicator method are equivalent here, they must rely on identical assumptions: mSITA and either CIT or CIO must hold in order for the missing indicator method to provide a consistent estimator of the ATE. 

In a slightly more complex scenario with one partially observed and one fully observed confounder ($C$), the propensity score for the MPA can be written as
\begin{equation*}
\begin{split}
logit(P(Z=1)) & = 
\left\{
\begin{array}{llr}
\alpha_1 + \beta_1X + & \gamma_1C & \text{ if } R=1 \\ 
\alpha_0 + {\color{white}\beta_1X +} & \gamma_0C & \text{ if } R=0. \\
\end{array}
\right.~\\
& =\alpha_0 + \beta_1X^*R + (\alpha_1 - \alpha_0)R + \gamma_0C +  (\gamma_1 - \gamma_0)CR.
\end{split}
\end{equation*} 
In contrast, the propensity score model for the missing indicator approach is:
\begin{equation*}
logit(P(Z=1)) = \alpha + \beta X^*R + \eta R + \gamma C.
\end{equation*}
This is the MPA model, constraining $\gamma_1$ to be equal to $\gamma_0$, i.e. the missing indicator model assumes the effect of the fully observed confounder on treatment is the same whether or not the other confounder is observed. 

In general, the missing indicator method is essentially a simplification of the MPA, which makes additional assumptions about the absence of interactions between the missingness indicator(s) and other fully observed confounders. These additional assumptions can be assessed in the data.

In summary, the missing indicator method relies on the same assumptions as the MPA, and additionally requires no effect modification of the fully observed confounder(s) by the missingness patterns.

\section{Plausibility of the CIT and CIO assumptions}\label{sec:investigatingMPAassmptns}
One of the CIT and CIO assumptions (equation \eqref{eq:CITandCIO}) must hold for the MPA to be valid. The plausibility of these assumptions in real-life settings will therefore determine how useful the MPA is as a missing data approach.

We have assumed that in the full data $X$ is a confounder, and so is associated with both treatment and outcome. The CIT assumption requires that the confounder-treatment relationship is absent in the subset of patients with $X$ unmeasured, whilst the CIO assumption requires that the confounder-outcome relationship is absent in patients with $X$ unmeasured. Thus, if either the CIT or CIO assumption holds, $X$ does not confound the relationship between treatment and outcome when it is missing (i.e.\ $X$ is not associated with both treatment and outcome in the subset of patients missing $X$). Informally, we refer to this property as $X$ being a confounder only when it is observed. 

The key point to consider is that the CIT and CIO assumptions are not about the missingness mechanisms that drive the missing data, as much as which relationships between variables exist in the subgroup of patients with missing confounder values.

\subsection{The CIT assumption: an illustrative example}\label{sec:CITokeg}
Consider a simplified version of our motivating example, investigating the effect of prescribing angiotensin-converting enzyme inhibitors or angiotensin receptor blockers (ACEI/ARBs) on the risk of acute kidney injury (AKI) using data from electronic health records. 

Underlying kidney function prior to ACEI/ARB prescription is a likely confounder: kidney function is a known risk factor for AKI and is likely to influence whether ACEI/ARBs are prescribed. Kidney function is classified into the stage of chronic kidney disease (CKD), via a serum creatinine blood test. 
Where a clinician ordered a kidney function test prior to the prescribing decision, it is reasonable to assume that the information regarding CKD stage contributed to that  decision. Where CKD stage was unavailable to the clinician, arguably it is unlikely to have influenced the prescribing decision. 

In a study relying on electronic health record data, the researcher has the information that was available to the clinician; CKD stage is missing in the research dataset when the clinician did not order the kidney function test. Thus it is plausible that treatment allocation is not associated with confounder information that is unavailable to the researcher.

In this simplified example, underlying CKD stage is always a risk factor for the outcome but is plausibly only associated with treatment allocation when it is measured. Thus, the CIT assumption holds; CKD stage is only a confounder when it is observed.

\subsection{The CIO assumption: an illustrative example}\label{sec:CIOokeg}
The CIO assumption is not plausible in the above simplified example (since underlying CKD stage, whether measured or unmeasured, is a risk factor for AKI). Indeed, in many clinical settings, this assumption would seem implausible. However, there are settings in which the CIO assumption may be plausible, such as the hypothetical example discussed below.

Suppose we were interested in estimating the effect of exposure to farming in early life on subsequent development of asthma. Childhood exposures to domestic allergens, such as dust mites, are potential confounders. Suppose that the relationship between dust mites and asthma has a threshold effect, i.e. an association is seen only once a certain concentration of dust mites is present. %https://www.nejm.org/doi/full/10.1056/nejm199008233230802

To address the research question, we might obtain routinely recorded information from health visitors regarding domestic exposures. Because health visitors do not collect information for the purposes of research, they might plausibly record information more thoroughly for households where there were concerns about the child's environment. Missing data for dust mites would therefore be more likely to occur in households with little evidence of dust mites, and less likely in households with a high concentration. 

In this example, we might expect concentration of dust mites to be associated with subsequent asthma only in households where dust mite concentration was recorded. In this case, CIO holds; dust mite concentration is a confounder only when measured.

\subsection{Summary}
We have described two simple scenarios in which it is plausible that either the CIT or the CIO assumption holds. A common feature of these two scenarios is that the studies described use routinely collected data. 

In the first, which uses electronic health record data to investigate effects of drug prescriptions, confounder data that is missing for the purposes of research was also absent when the prescribing decision was being made. In such cases, it seems plausible that the the missing information, which was unavailable to the clinician, did not contribute to their prescribing decision. Thus electronic health record research, investigating effects of drug  prescriptions, may be an area where the MPA is likely to prove useful.  

The second scenario above describes a situation where the data collection process itself depends on whether or not the patient belongs to the missingness pattern where the confounder-outcome association exists; confounder data are missing on an individual precisely because the health worker is satisfied that the individual's value of the confounder is low enough that it would not affect their health. In these scenarios, the CIO assumption may be plausible. 

\section{Detecting and dealing with violations of the MPA assumptions}\label{sec:method}

The mSITA assumption (equation \eqref{eq:mSITA}), and the CIO and CIT assumptions (equation \eqref{eq:CITandCIO}) are statements of conditional independence. In this section, we describe how causal diagrams can be used to assess conditional independence statements. We demonstrate the use of causal diagrams in a simple scenario in order to draw some general conclusions about situations in which the MPA's assumptions are likely to be violated. We also use causal diagrams to explore when the MPA's assumptions are violated in a range of simple settings and which violations we can `handle', i.e.\ which violations can be removed by measuring factors which affect whether or not the confounder is missing.

\subsection{Causal diagrams}
Causal diagrams, or directed acyclic graphs, are a useful tool for assessing conditional independencies under an assumed causal structure. Nodes represent variables, whether measured or unmeasured. Arrows between nodes represent causal relationships. 

Because the assumptions of the MPA involve the potential, rather than observed, outcomes we turn to a specific type of causal diagram: Single World Intervention Templates (SWITs) \cite{SWIGs}.

\subsubsection{Single World Intervention Templates}
SWITs are standard directed acyclic graphs which have been adapted to show potential, instead of observed, outcomes. This involves `splitting' the treatment node into two halves; the first represents the observed treatment $Z$, and the second represents an `intervened-on value', $z$. Determinants of observed treatment affect the first half (i.e. incoming arrows go into the $Z$ half), and effects of treatment are determined by the second (i.e. outcoming arrows leave from the $z$ half). A consequence of this splitting is that variables affected by treatment now become potential rather than observed variables.

Figure \ref{fig:dagintro} shows a simple SWIT representing a typical confounding scenario where the confounder $X$ has a causal effect on the treatment and the outcome. Additionally, this graph encodes the assumption that the missingness indicator $R$ (i.e. whether or not the confounder is missing) is associated with the treatment and the outcome, via shared common causes in both cases (denoted $U_Z$ and $U_Y$ respectively). In Figure \ref{fig:dagintro}, just the outcome is affected by treatment and thus this SWIT includes the potential outcome $Y(z)$ rather than the observed outcome $Y$.

\subsubsection{D-separation}
A rule called d-separation, proposed within the context of directed acyclic graphs \cite{Pea95} and extended to SWITs \cite{SWIGs}, determines whether a particular conditional dependency holds or not, under an assumed causal structure. Broadly speaking, association is transmitted through series of arrows --- paths --- in the assumed causal diagram \cite{Pea09}. A particular path will transmit association between the nodes at either end unless it contains a `collider':  a node which  --- in that path --- has two incoming arrows. In Figure \ref{fig:dagintro}, the path $Z \leftarrow X \rightarrow Y(z)$ will transmit association between $Z$ and $Y(z)$, but the path $Z\leftarrow U_Z \rightarrow R \leftarrow U_Y \rightarrow Y(z)$ will not because R is a collider in this path. Conditioning on a non-collider blocks associations through a specific path. Conversely, conditioning on a collider removes the blockage through that collider thereby allowing association to be transmitted. Introducing bias by conditioning on a collider is often termed collider bias \cite{Gre03}.

The d-separation rule states that two variables in the assumed causal diagram are conditionally independent given a set of variables $V$ if for each path connecting the two variables: (i) the path contains two arrows which collide at a node in the path, and that node is neither in $V$, nor a cause of a variable in $V$; or (ii) the path has a non-collider which is in $V$ \cite{Pea95,Pea09}.  

In Figure \ref{fig:dagintro}, there are two paths between $Z$ and $Y(z)$: $(Z\leftarrow X\rightarrow Y(z)$, and $Z\leftarrow U_Z\rightarrow R\leftarrow U_Y\rightarrow Y(z))$. If the conditioning set is $V=\{X\}$, then $Z$ and $Y(z)$ are conditionally independent given $V$. This is because the first path contains a non-collider ($X$) which is in $V$ (condition (ii)) and the second contains a collider ($R$) which is not in V (condition (i)). In contrast, $Z$ and $Y(z)$ are not conditionally independent given $V=\{X, R\}$, because the second path then contains a collider (i.e.\ $R$) which is in $V$, and neither $X$ nor $R$ is a non-collider in this path.

\subsection{Assessing the MPA's assumptions}
In this subsection, we apply d-separation to Figure \ref{fig:dagintro} to assess whether the mSITA, CIT and CIO assumptions hold in this scenario.

\subsubsection{Assessing the mSITA assumption}
Suppose Figure \ref{fig:dagintro} depicts the true underlying causal structure which gave rise to our study data. With a single partially observed confounder, the mSITA assumption states that $Z \perp Y(z) | X, R$. By applying d-separation to Figure \ref{fig:dagintro} as described above, we find that $Z$ is not conditionally independent of $Y(z)$ given $X$ and $R$ in Figure \ref{fig:dagintro}; the mSITA assumption is violated in this scenario. 

For more complex causal diagrams, it may help to use software such as Dagitty to assess which conditional independencies hold \cite{DAGitty}. R code which uses Dagitty to check the MPA's assumptions for the scenario shown in Figure \ref{fig:dagintro} can be found in Appendix \ref{sec:appx:dagitty}.

\subsubsection{Assessing the CIT/CIO assumptions}

The CIT and CIO assumptions state that $Z \perp X_{mis} | X_{obs}, R$, and $Y(z) \perp X_{miss} | X_{obs}, R$, respectively. With a single confounder $X$, these assumptions are trivially true in the subgroup of patients with $X$ observed (because $X_{mis}$ is empty given $R=1$). In the subgroup of patients with $X$ missing, the assumptions become: $Z \perp X | R=0$, and $Y(z) \perp X | R=0$, respectively. 

\begin{figure}[h]
	\begin{centering}
		\includegraphics[width=0.4\textwidth]{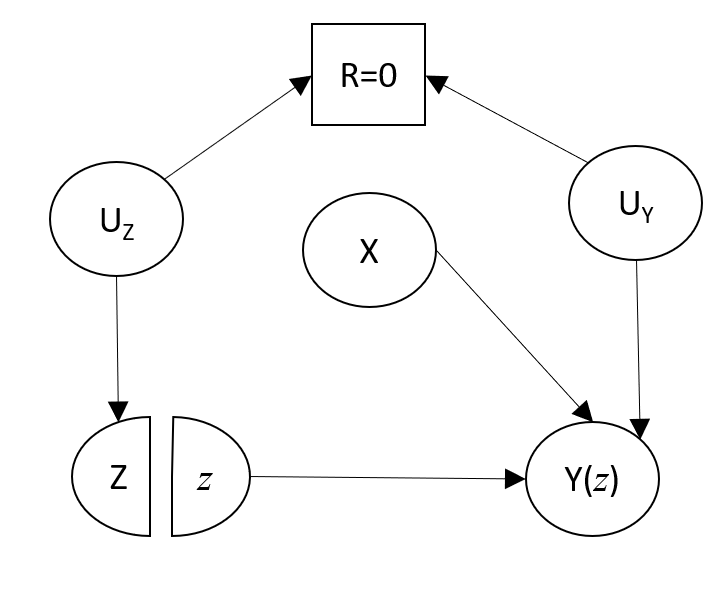}
		\caption{Simple causal diagram modified to assess the CIT/CIO assumptions. \newline { $X$: confounder. $Z$: treatment. $Y(z)$: potential outcome resulting from intervening to set $Z$ equal to a particular value $z$. $R$: missingness indicator (=1 if $X$ observed, =0 if $X$ is missing). $U_Z$: unobserved common cause between $R$ and $Z$. $U_Y$: unobserved common cause between $R$ and $Y(z)$.}}\label{fig:DAGcitcio}
	\end{centering}
\end{figure}

A minimum condition for CIT or CIO to be satisfied is that $X$ cannot be a confounder when it is missing. Thus, for either of these assumptions to hold, there must be grounds for believing that the causal relationships that generate confounding bias in the full data are different in the subgroup with missing confounder values (compared to the subgroup with observed confounder values). For example, in Figure \ref{fig:dagintro}, if we believe that all the arrows shown exist in the subgroup with missing confounder values, then both CIT and CIO would be violated. In contrast, suppose we believe that this diagram depicted the correct situation with full data, but we believe that the arrow from the confounder to treatment did not exist when $X$ was missing (i.e. when $X$ is unmeasured it does not contribute to the treatment allocation decision; see Section \ref{sec:CITokeg} for an example where this is plausible). In this case,  Figure \ref{fig:DAGcitcio} would depict the underlying causal structure for the subgroup with $X$ unmeasured. The square box around $R$ denotes the restriction of our attention to this subgroup. 

In Figure \ref{fig:DAGcitcio}, the only path connecting $Z$ and $X$ passes through $Y(z)$, a collider on the path; thus applying the d-separation rule shows that $Z$ and $X$ are conditionally independent (in the subgroup with $R=0$). Here, the CIT assumption holds. Because there is a direct arrow from $X$ to $Y(z)$, however, CIO does not hold.

\subsection{Key violations of the MPA's assumptions}

In this section, we use causal diagrams to explore when the MPA's assumptions are violated in a range of simple settings.

\textbf{ Scenarios considered:} We consider scenarios where the outcome $Y$ and treatment $Z$ are fully observed. Initially, we focus on simple scenarios with a single partially observed confounder, $X$. Subsequently we extend this to consider scenarios with an additional, fully observed confounder, $C$. We consider all combinations of the scenarios discussed below, omitting those which give rise to cycles (i.e. we do not allow scenarios where a variable has a causal effect on itself). 

\textbf{ Relationships between the confounder, treatment, and outcome:} We consider causal structures where the relationships between the confounder $X$ and the treatment and outcome are either a direct causal relationship (e.g. $X$ causes treatment), or via shared unmeasured common causes (e.g. a third factor causes both $X$ and treatment). The relationship between the confounder and the treatment is allowed to differ depending on whether the confounder is observed or missing; specifically, this relationship is allowed to be absent when $R=0$.  Similarly, the presence or absence of the relationship between the confounder and outcome is allowed to depend on $R$. This allows for $X$ to be a confounder only when observed, as discussed in the previous section. 

\textbf{ Missingness mechanisms:} For each of the confounder, treatment and outcome, we considered: no relationship with the missingness indicator, a causal effect on the missingness indicator, the missingness indicator has a causal effect on the variable, or an unobserved common cause with the missingness indicator (allowing scenarios where one or more variables have both a direct causal relationship and a common cause with the missingness indicator). 

When a variable has a causal effect on the outcome, we assume that this effect operates on the potential outcome rather than the observed (e.g. $X$ causes $Y(z)$ rather than $X$ causing $Y$). Conversely, in the case where outcome is a cause of missingness, we have chosen to allow the observed outcome to cause missingness ($Y$ causes $R$) rather than the potential outcome, since this is arguably more plausible in real data.

\textbf{ Assessment of assumptions:} In each setting, we draw the appropriate causal diagram and assess the assumptions by applying d-separation to the causal diagram overall, and to the modified causal diagram restricted to the subgroup with $X$ missing. 

In some scenarios, a slightly more complex route must be taken to assess the conditional independencies involved in the MPA assumptions. If the treatment or outcome is a cause of missingness then the relevant SWIT contains $R(z)$, the `potential' missingness after intervening on treatment, rather than the observed pattern of missingness. Thus we can no longer use this graph to assess the relevant assumptions. In these cases we turn to twin networks \cite{Bal94,Shp07} (Appendix \ref{sec:appx:TN}).

\subsubsection{Key violations of the mSITA assumption}

In the scenarios we considered, most violations of mSITA occurred via collider bias on $R$. In order for this type of violation to occur, there needs to be a path from $Z$ to $R$ \emph{and} a path from $Y(z)$ to $R$, each ending with arrows pointing towards $R$. These violations operate via a cause of $R$. We let $U_X$ represent common causes of missingness and the confounder, $U_Z$ represent common causes of missingness and the treatment, and $U_Y$ represent common causes of missingness and the outcome.

\begin{figure}[h]
	\begin{centering}
		\includegraphics[width=1.0\textwidth]{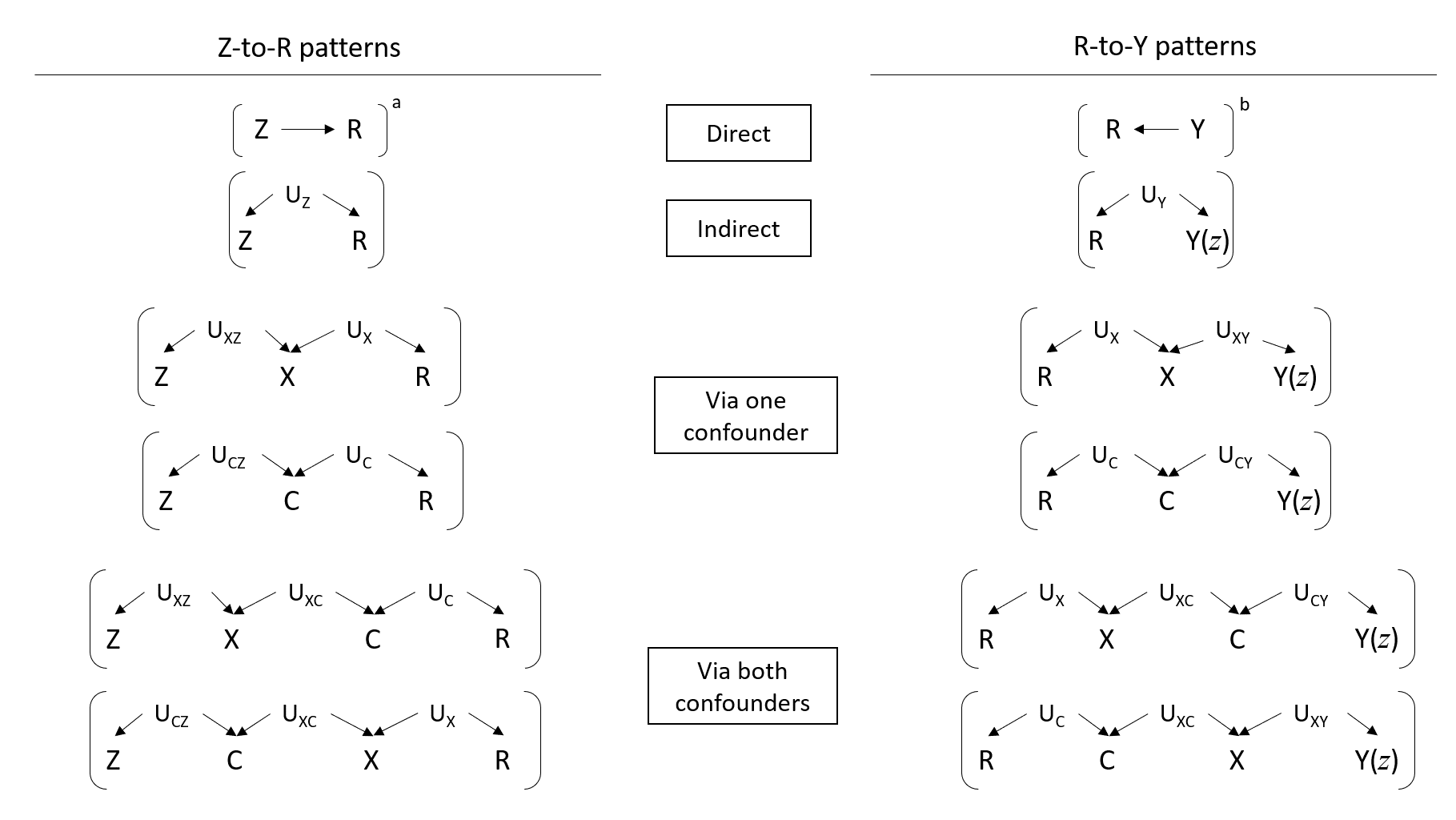}
		\caption{Summary of violations of the mSITA assumption. If one of the `Z-to-R' patterns and one of the `R-to-Y' patterns occurs in the causal diagram representing the study in question then the mSITA assumption will be violated. \newline { $^a$ Also a violation if this occurs with additional `R-to-Y patterns' shown in Appendix \ref{sec:appx:assumptions_extra}; $^b$ Sufficient condition on its own, without a `Z-to-R pattern'.\newline $X$: partially observed confounder. $C$: fully observed confounder. $Z$: treatment. $Y(z)$: potential outcome resulting from intervening to set $Z$ equal to a particular value $z$. $Y$: observed outcome. $R$: missingness indicator (=1 if $X$ observed, =0 if $X$ is missing). $U_{st}$: unobserved common cause between two variables $s$ and $t$. $U_{s}$: unobserved common cause between $R$ and another variable $s$. }}\label{fig:msita}
	\end{centering}
\end{figure}

The different `Z-to-R' and `R-to-Y' patterns that could occur are summarised in Figure \ref{fig:msita}. If one (or more) of each of these two patterns occurs then mSITA will be violated. For example, Figure \ref{fig:dagintro} shows the violation which arises when both the indirect `Z-to-R' pattern and the indirect `R-to-Y' pattern occur (both patterns in the second row of Figure \ref{fig:msita}).

A key result in Figure \ref{fig:msita}  is that when treatment and missingness are associated via shared common causes, and outcome and missingness are associated via shared common causes, then mSITA is violated (as shown in Figure \ref{fig:dagintro}). So the MPA cannot be used in scenarios where there are unmeasured determinants of confounder missingness which are also associated with the treatment and the potential outcomes. 

Another important result in Figure \ref{fig:msita}  is that if the outcome has a causal effect on confounder missingness, i.e. if $Y \rightarrow R$, then mSITA is violated without the need for any `Z-to-R' patterns. So the MPA cannot be used in scenarios where outcome affects whether or not confounder values are missing. For example, in our AKI example, suppose that more efforts were made to track down historical laboratory measures of eGFR for patients who were diagnosed with AKI, then this would immediately violate the mSITA assumption. 

A third important result is that when treatment causes missingness, and missingness in turn has a causal effect on the potential outcomes, mSITA is violated (see footnote a in Figure \ref{fig:msita}), although whether this is likely to occur in practice is unclear.

\subsubsection{Handling violations of the mSITA assumption}
All violations of mSITA, other than those involving the treatment or the outcome causing missingness of the confounder, operate via a cause of $R$. Suppose it were possible to measure such variables with no missingness. We could define a new set of confounders $\tilde{X}=\{X, U_X, U_Z, U_Y\}$ (or, where there is an additional fully observed confounder $C$, $\tilde{X}=\{X, C, U_X, U_C, U_Z, U_Y\}$). Including this new set of confounders in the propensity score model, and thus the conditioning set for the mSITA assumption, removes the violation of this assumption. In most cases, measuring a subset of these variables will suffice. For example, in  Figure \ref{fig:dagintro}, if $U_Z$ could be measured and included in the propensity score model, the mSITA assumption would become: $Z \perp Y(z) | X, R, U_Z$, which is satisfied in Figure \ref{fig:dagintro}.

\subsubsection{Key violations of the CIT and CIO assumptions}
\begin{figure}[h]
	\begin{centering}
		\includegraphics[width=1\textwidth]{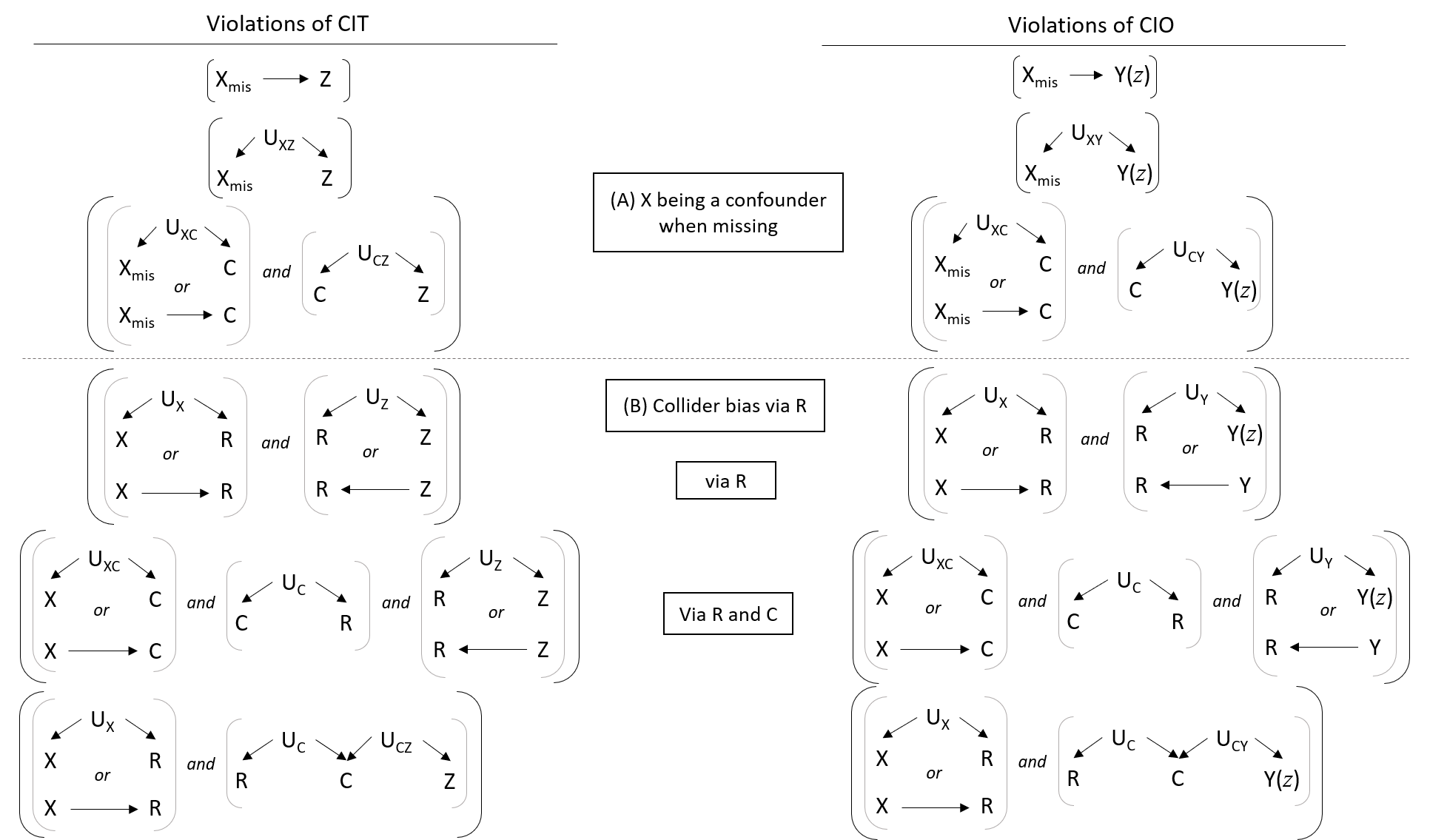}
		\caption{Summary of violations of the CIT and CIO assumptions. If one or more of the six sets of conditions on the left hand side appear in the relevant causal diagram (modified to reflect relationships in the subgroup with $X$ unobserved i.e.\ restricted to $R=0$), the CIT is violated. Similarly, if any of the six sets of conditions on the right hand side occur then CIO is violated. Additional violations involving $Y \rightarrow R$ and $Z \rightarrow R$ can be found in Appendix \ref{sec:appx:assumptions_extra}. \newline { $X$: partially observed confounder. $C$: fully observed confounder. $Z$: treatment. $Y(z)$: potential outcome resulting from intervening to set $Z$ equal to a particular value $z$. $Y$: observed outcome. $R$: missingness indicator (=1 if $X$ observed, =0 if $X$ is missing). $U_{st}$: unobserved common cause between two variables $s$ and $t$. $U_{s}$: unobserved common cause between $R$ and another variable $s$. }}\label{fig:citcio} 
	\end{centering}
\end{figure}

Figure \ref{fig:citcio} summarises the possible violations of CIT and CIO, which fall into two broad groups: (A) violations related to $X$ being a confounder when it is missing, and (B) violations due to collider bias via $R$. 

Since mSITA is always violated if outcome causes missingness, some CIT/CIO violations involving $Y \rightarrow R$ are shown only in Appendix \ref{sec:appx:assumptions_extra}, along with a few additional violations involving $Z \rightarrow R$. 

Group (A) violations in Figure \ref{fig:citcio} relate to $X$ being a confounder only when observed, in the sense that if one of the CIT group (A) violations or one of the CIO group (A) violations held, $X$ would be a confounder when missing. For these violations, $X$ has been replaced by $X_{mis}$ to emphasise the fact that we need to focus on relationships that exist in the subgroup of patients with a missing confounder value when assessing this assumption. 

In contrast, Group (B) violations relate to collider bias induced by conditioning on $R$.

\subsubsection{Handling violations of the CIT and CIO assumptions}
As with violations of the mSITA assumption, many of the violations of the CIT and CIO assumptions --- specifically those belonging to Group (B) --- can be removed by measuring and conditioning on causes of $R$, i.e. factors which determine whether or not the confounder is measured. 

One notable exception, in which no additional conditioning can prevent the violation of these assumptions, is scenarios where both the confounder and the treatment cause the missingness, or both the confounder and outcome cause the missingness. In the former case, the CIT assumption is violated; in the latter, CIO (and mSITA) is violated.

\subsection{Summary}\label{sec:violconcl}
We have shown how causal diagrams can be used to assess the plausibility of the MPA's assumptions and, using this, we have identified causal structures that lead to violations of these assumptions.

The key assumption required by the MPA is that the confounder acts as a confounder only when observed. Thus for the MPA to be an appropriate method to use, we must believe that the relationships between treatment, outcome, and confounder are different in the subgroup with the confounder unmeasured. 

If this is a plausible assumption, care must be taken to ensure that additional violations of the mSITA, CIT and CIO assumptions do not arise from collider bias via missingness. Generally, measuring all factors which determine whether or not the confounder is measured, and including those factors in the propensity score model, will remove such violations of the assumptions, although identifying and measuring such factors may be difficult in practice. 

A few scenarios are of particular note. The MPA's assumptions do not hold if: (I) outcome affects missingness of the confounder; (II) outcome and missingness have shared unmeasured common causes, and treatment and missingness have shared common causes; or (III) the confounder and treatment both affect missingness of the confounder and the confounder is associated with outcome in the subgroup with $X$ missing. These scenarios  merit careful consideration to ensure the MPA's appropriateness in a given clinical setting.

\section{Practical guide to assessing the mSITA, CIT and CIO assumptions}\label{sec:practical_guidance}

In order to decide if the MPA's assumptions hold in a particular clinical setting, the first, most important step, is to assess whether it is plausible for the partially observed confounder to be a confounder only when observed. In other words, is there any reason to believe the relationships between treatment, outcome and confounder are different in the subset of patients with missing confounder values.

Second, the key scenarios (I), (II) and (III) noted at the end of the previous section, in which the MPA's assumptions do not hold, should be carefully considered using substantive knowledge to ensure these do not apply in the setting at hand. 

Third, a causal diagram should be constructed, reflecting what is believed to be the underlying clinical structure. As with any causal diagram, any variable --- measured or unmeasured --- which may have a causal effect on two or more variables in the causal diagram must also be included. Missingness indicators for the partially observed confounders should be included in the causal diagram at this stage. When there are multiple partially observed confounders, the causal diagram will include one missingness indicator per partially observed confounder. 

Fourth, the causal diagram should be converted into a SWIT or a twin network, as appropriate. Once the SWIT or twin network has been created, d-separation can be applied to determine whether the mSITA assumption holds. Appendix \ref{sec:appx:dagitty} provides R code to do this for Figure \ref{fig:dagintro}, and for the causal diagram associated with our more complex motivating example. 

To assess CIT and CIO, the SWIT or twin network should be modified to reflect the relationship thought to be absent in the subgroup of patients with missing confounder values (i.e. remove the arrows which reflect the assumption that the confounder is only a confounder when observed). In this modified diagram, the d-separation rule can be again applied to assess CIT and CIO.

\subsection{Assessing the validity of the assumptions in the motivating example}

\subsubsection{Confounders only when observed}
For the MPA to be appropriate in this example, we have to believe that the two partially-missing confounders --- ethnicity and baseline chronic kidney disease (CKD) stage --- act as confounders only when observed. If baseline CKD stage is not available, this unobserved information cannot be used to determine the General Practitioner's treatment decision whether or not to prescribe ACEI/ARBs. In practice, CKD stage may be recorded in a part of the patient record that the General Practitioner is aware of but researchers using CPRD data cannot access (e.g.\ letters from secondary care). However, this is likely to reflect advanced CKD for only a small proportion of the whole study population. So in general, it seems plausible that baseline CKD stage affects the clinician's prescribing decision only when recorded. If the clinician believes that the patient's ethnicity is an important reason for prescribing, or not, ACEI/ARBs, then it is likely that they would ensure this information is recorded. Thus it is plausible that ethnicity affects treatment allocation only when it is measured.

Therefore, the CIT assumption may be reasonable for this scenario. Conversely, both baseline CKD stage and ethnicity are risk factors for AKI, whether measured or not. Thus the CIO assumption is not plausible here.

\subsubsection{Checking plausibility of key violations}
For the MPA to be appropriate in this example, we also need grounds to believe that the three key scenarios mentioned in Section \ref{sec:violconcl} do not apply in this setting.

Scenarios (I) and (III) rely on either outcome or treatment affecting missingness of the confounders. As CKD stage was defined at baseline, missingness of baseline CKD stage precedes treatment and, as a result, outcome. It also seems plausible that missingness of ethnicity occurs prior to treatment and outcome. Hence we believe that these scenarios do not apply here.

Scenario (II) is when outcome and missingness have shared unmeasured common causes, and treatment and missingness have shared common causes. Baseline CKD stage is more likely to be recorded for patients expected to have a higher risk of kidney disease due to age or chronic comorbidities (eg.\ hypertension, diabetes) or due to other signs that the patient has poor kidney function (i.e.\ CKD itself may affect the chance of the clinician measuring CKD stage). Whilst these risk factors are associated with missingness and treatment or outcome, they are already captured in the electronic health records. 

With respect to ethnicity, patients who are hospitalised are more likely to have ethnicity recorded (due to linkage of primary and secondary care data). Missingness of ethnicity may be caused by service-level factors such as level of administrative support at the time patients are admitted to hospital. It seems unlikely that these factors are also determinants of treatments previously prescribed in primary care or whether patients develop acute illnesses that require admission to hospital. Since we believe that any relevant common causes are measured, scenario (II) does not apply in our setting.

After considering the three key scenarios mentioned in Section \ref{sec:violconcl} in which the MPA's assumptions do not hold, we have found that these do not seem plausible in our motivating example. Having ruled out these violations, we proceed to the next step of our framework: to develop a causal diagram.

\subsubsection{Developing a causal diagram}
Figure \ref{fig:akiexample} shows the causal diagram developed for this example. This diagram encodes the investigators' assumptions that age, sex and ethnicity each affect both treatment and outcome. Age and sex affect the risk of developing diabetes,  CKD, ischaemic heart disease, cardiac failure, arrhythmia and hypertension. 

Figure \ref{fig:akiexample} is a SWIT, i.e.\ the treatment node, representing prescription of ACEI/ARBs, has been split into two: `Ace' and `ace', with the former representing the observed treatment and the latter representing the intervened-on treatment. Thus patient factors affect `Ace' but not `ace', and only `ace' affects subsequent AKI.

\subsubsection{Assessing the mSITA assumption}

The mSITA assumption, for the motivating example, says that:  $$Z \perp Y(z) |  R_{ckd}, R_{eth}, Ckd, Eth, V,$$ where $Z$ represents ACEI/ARB prescription; $Y(z)$ the potential outcome (AKI status that would be observed if the patient were prescribed level $z$ of ACEI/ARB); $V$ represents the confounders age, sex, diabetes, ischaemic heart disease, cardiac failure, arrhythmia and hypertension, $R_{eth}$ and $R_{ckd}$ are missingness indicators for ethnicity and baseline CKD stage, and $Ckd$ and $Eth$ are the confounders CKD stage and ethnicity, respectively. 

D-separation can be applied to the SWIT in Figure \ref{fig:akiexample} manually, to assess whether this conditional independence holds under the causal assumptions encoded in the diagram. Alternatively, Appendix \ref{sec:appx:dagitty_aki} contains R code to use Dagitty to assess the mSITA assumption. In this case, the conditional independence statement is true; mSITA holds under the assumed causal diagram.

\subsubsection{Assessing the CIT and CIO assumptions}
We have already established that CIO does not hold in our motivating example. The CIT assumption states that:
\begin{align*}
Z \perp (Ckd, Eth) & \ | \  R_{ckd}=0, R_{eth}=0, \qquad  V, \\
Z \qquad \perp Ckd & \ | \  R_{ckd}=0, R_{eth}=1, Eth, V, \\
Z \qquad  \perp Eth & \ | \  R_{ckd}=1, R_{eth}=0, Ckd, V. 
\end{align*}
To assess the first of these, we create a modified version of Figure \ref{fig:akiexample} which omits the arrows that we do not think exist when both ethnicity and baseline CKD stage are missing. So we remove the arrow from baseline CKD stage to ACEI/ARB prescription, and we remove the arrow from ethnicity to ACEI/ARB prescription.  

In this modified diagram, we then assess whether, after conditioning on the two missingness indicators and the fully observed confounders, the treatment is independent of both ethnicity and baseline CKD stage.  This is done by applying d-separation for each partially observed confounder, either manually or using Dagitty via R (code in Appendix \ref{sec:appx:dagitty_aki}). In this case, the conditional independence holds. 

This process is repeated in the two subgroups where only one of ethnicity and baseline CKD stage is recorded, assessing the second and third independence statements above in the appropriately-modified causal diagrams. In each case, the relevant conditional independence holds.

Thus, under the assumed causal diagram, the CIT assumption holds.

\subsubsection{Conclusion}
If our causal diagram correctly represents the causal structure giving rise to our study data, both mSITA and CIT hold. Under these two assumptions the MPA will provide consistent estimates of the ATE. 

\section{Motivating example: applying the MPA} \label{sec:results_eg}
\subsection{Methods}
We estimated the effect of prescription of ACEI/ARBs on the incidence of AKI within 5 years of follow-up as a risk difference, first with no adjustment for confounding, and then by using inverse probability of treatment weighting (IPTW). For IPTW, we estimated propensity scores using logistic regression to model ACEI/ARB prescription as a function of the covariates: age, sex, baseline CKD stage, ethnicity, diabetes mellitus, ischaemic heart disease, arrhythmia, cardiac failure and hypertension (including an interaction between age and ischaemic heart disease, and an interaction between age and hypertension). We applied non-parametric bootstrapping (500 replications of the combined process of propensity score estimation and treatment effect estimation) to obtain Normal approximation 95\% confidence intervals.

To deal with missing data in baseline CKD stage and ethnicity when estimating the propensity score, we applied complete records analysis, the MPA, the missing indicator approach and multiple imputation. For the MPA,  the propensity scores was estimated separately in the four subgroups corresponding to whether or not baseline CKD stage and ethnicity were measured. For the missing indicator approach, we added `missing' categories to each of baseline CKD stage and ethnicity. For multiple imputation, 10 imputed datasets were created using chained equations. The imputation model included AKI incidence within 5 years, ACEI/ARB prescription and all covariates and interactions included in the propensity score model. In each imputed dataset, propensity scores were estimated and IPTW was used to obtain treatment effect estimates, which were then pooled using Rubin's rules \cite{Ley17}. Standardized differences (see \cite{Wil14introps} for details) were calculated in the original sample to assess covariate balance, and after IPTW with each of the analysis methods used.

\subsection{Results}
The complete records analysis included $121,527$ patients with full data. All other missing data methods included all $570,586$ patients. Using any of the analysis methods with IPTW removes most of the imbalance present in the original dataset (Table \ref{tab:akibal} in Appendix \ref{sec:appx:covbalance}). Estimates of the effect of ACEI/ARBs on AKI are shown in Table \ref{tab:akires}. All missing data methods greatly reduce the crude estimate of effect, with complete records analysis providing the smallest estimate and multiple imputation providing the estimate closest to the crude analysis. The MPA and missing indicator method produce almost identical results, estimating that patients prescribed ACEI/ARBs had 6 additional cases of AKI within 5 years, per 1000 people, with a 95\% confidence interval of (5,7), compared to patients who were not prescribed ACEI/ARBs. 

In terms of precision, the complete records analysis has a very wide confidence interval, in contrast to the other missing data methods which all produce much narrower confidence intervals.

\subsection{Conclusion}
In this example, the mSITA and CIT assumptions both appear to be plausible. Under these assumptions we can expect the MPA to provide a consistent estimate of the treatment effect.  

The assumption that data are missing at random (MAR) is questionable in this example as baseline CKD stage is more likely to be recorded for patients with a lower level of kidney function (e.g.\ if they are ill or have more risk factors for kidney disease that have led to testing) \cite{McD16} and therefore baseline CKD stage might be missing not at random. The MPA's assumptions appear plausible and the approach does not rely on the MAR assumption. However, multiple imputation and the MPA provide fairly similar estimates in this example. A possible explanation is that departures from the MAR assumption may be weak since we have incorporated factors related to a lower level of kidney function (i.e.\ assumptions for multiple imputation may be approximately satisfied).

However, since factors related to a lower level of kidney function are likely already captured in the observed data, the departure from the missing at random assumption may be small. This may explain why multiple imputation and the MPA provide fairly similar estimates in this example, with imputation giving an estimate closer to the crude estimate.

Multiple imputation, as typically applied, relies on the assumption that data are missing at random i.e.\ that missingness can be explained by the observed data \cite{Car13,Ste09}. Here, the assumption that data are missing at random is questionable as baseline CKD stage is more likely to be recorded for patients with a lower level of kidney function (e.g.\ if they are ill or have more risk factors for kidney disease that have led to testing) \cite{McD16} and therefore baseline CKD stage might be missing not at random.  The MPA offers an alternative which does not rely on the assumption of MAR. Rather, the validity of the MPA depends on the underlying causal structure being different in different missingness patterns. That being said, imputation and the MPA provide fairly similar estimates in this example, although imputation gives an estimate closer to the crude estimate.

Analysing only complete records potentially leads to bias and results in a loss of efficiency: indeed, the complete records estimate has low precision, due to the exclusion of a large portion of the data. This precision is recovered by the MPA, the missing indicator approach and multiple imputation. 

\section{Discussion}
\label{sec:discussion}

We have explored the three assumptions underlying the missingness pattern approach to dealing with missing counfounders in propensity score analysis. We have described how d-separation can be applied to a causal diagram to assess the MPA's assumption in a given setting and provided a framework and detailed example to allow researchers to ensure the appropriateness of this method in practice.

The key assumption required by the MPA is that the confounder acts as a confounder only when observed. Thus for the MPA to be an appropriate method to use, we must believe that the relationships between treatment, outcome, and confounder are different in the subgroup with the confounder unmeasured. While this assumption will be plausible only in specific scenarios, one setting where it may have broad applicability is in the area of electronic health record research. In such studies, missing confounder information reflects information that the clinician did not have when making prescribing decisions, thus the assumption that the missing values did not affect prescribing may well be reasonable.

If this key assumption is thought to be satisfied, careful consideration is required to ensure that the remaining assumptions of the MPA are satisfied. In particular, the assumptions do not hold in the following scenarios: (i) where the outcome affects missingness of the confounder; (ii) where outcome and missingness have shared unmeasured common causes, and outcome and treatment have shared common causes; and (iii) where a partially-missing confounder and treatment both affect missingness of the confounder and the confounder is thought to be associated with outcome whether or not it is measured.

We also found that many violations of the MPA's assumptions can be dealt with by recording, and including in the analysis, factors which affect whether the confounders are missing or not. Thus, although measuring such factors may be difficult in practice, careful consideration of the process by which data become missing is essential. 

Our results demonstrate that classification of the missingness mechanism according to Rubin's taxonomy does not provide information as to whether the MPA's assumptions will hold. Unlike most missing data methods, for example, data being missing completely at random does not guarantee that the assumptions of the MPA are satisfied: the underlying relationships between the partially missing confounder and either the treatment or outcome (or both) would need to differ according to whether or not the confounder was missing. Also, if a confounder is missing not at random, but the confounder does not affect treatment when missing, the MPA's assumptions may hold.

The missing indicator method is a popular and easy method to deal with missing confounder data \cite{Gre95,Gro12}. However, it is believed to be an `ad hoc' method \cite{Gre95} that produces biased results \cite{Gro12}. Although the missing indicator method is indeed biased under standard missing at random assumptions \cite{Gro12}, our results show that in the propensity score context, the missing indicator approach is a simplified version of the MPA, and hence requires the same assumptions for valid results, along with additional assumptions about interaction terms in the propensity score model.  Our work, therefore, allows researchers to use the missing indicator approach in a principled way.

There are several advantages to using the MPA, or the simpler missing indicator method, when dealing with partially observed confounders in propensity score analysis. First, the method itself is simple to comprehend and easy to implement. Second, in contrast to complete records analysis, the MPA retains all patients in the analysis. Third, the MPA may be appropriate in some situations where multiple imputation is not, as the MPA does not require the missing at random assumption to hold.

A limitation of the MPA is that we require sufficient sample size in each missingness pattern in order to be able to estimate propensity scores. This is of particular concern when there are many missingness patterns, a scenario to which the MPA is not currently easily extendable. Qu and Lipkovich (2009) suggested a pattern pooling algorithm \cite{Qu09} to ensure sufficient sample size when estimating propensity scores when there are a large number of missingness patterns. Further work is needed to explore the performance of their algorithm in a range of scenarios. An extension to the MPA was proposed by D'Agostino et al. (2001) \cite{DAg01}. They suggested that in each missingness pattern, propensity scores should be estimated in the wider group of all subjects with observed data for the relevant confounders, retaining estimated propensity scores only for those who actually observed that particular pattern. Further work is required to compare this extension with the original MPA, and to investigate how to account for the correlation induced by this method.

We have concentrated on scenarios where treatment and outcome are both fully observed. A hybrid method, combining the MPA and multiple imputation, was proposed by Qu and Lipkovich (2009) \cite{Qu09} and studied by Seaman and White (2014) \cite{Sea14}.

The MPA is simple and easy to implement, and may be useful in settings where other missing confounder data methods are not appropriate. We believe that this approach will be particularly useful in areas using routinely collected data, particularly electronic health record research. We have produced practical guidance for researchers to decide whether the underlying assumptions of the MPA are plausibly satisfied in a particular clinical setting. 

% The Acknowledgements section must be placed just before the References.
\section*{Acknowledgements}
This work was supported by the Economic and Social Research Council [Grant Number ES/J5000/21/1]; the Medical Research Council [Project Grant MR/M013278/1]; and by Health Data Research UK [Grant Number EPNCZO90], which is funded by the UK Medical Research Council, Engineering and Physical Sciences Research Council, Economic and Social Research Council, Department of Health and Social Care (England), Chief Scientist Office of the Scottish Government Health and Social Care Directorates, Health and Social Care Research and Development Division (Welsh Government), Public Health Agency (Northern Ireland), British Heart Foundation and Wellcome. Ethics approval was given by the London School of Hygiene and Tropical Medicine Research Ethics Committee [Reference: 15880] and by the Clinical Practice Research Datalink Independent Scientific Advisory Committee [ISAC Protocol Number 14\_208A2].

\clearpage
% Tables must be on separate pages after the reference list
\newpage
\begin{table}[htbp]
	\centering
	\caption{Patient characteristics by prescription of ACEI/ARBs}\label{tab:akichar}
	\footnotesize
	\begin{tabular}{rrrrrrrr}
		\hline
		\multicolumn{2}{c}{Baseline} &       & \multicolumn{5}{c}{Prescribed ACEI/ARB} \\
		\multicolumn{2}{c}{Characteristic} &       & \multicolumn{2}{c}{Yes (n (\%))} &       & \multicolumn{2}{c}{No (n (\%))} \\
		& \multicolumn{2}{c}{ } & \multicolumn{2}{c}{(Total = 159,389)} &       & \multicolumn{2}{c}{(Total = 411,197)} \\
		\hline
		Age (years) & 18 to 42 &       & 16,616 & (10.4\%) &       & 94,265 & (22.9\%) \\
		& 43 to 53 &       & 39,541 & (24.8\%) &       & 77,224 & (18.8\%) \\
		& 54 to 62 &       & 36,325 & (22.8\%) &       & 77,985 & (19.0\%) \\
		& 63 to 71 &       & 30,667 & (19.2\%) &       & 75,141 & (18.3\%) \\
		& $\geq$ 72 &       & 36,240 & (22.7\%) &       & 86,582 & (21.1\%) \\
		&       &       &       &       &       &       &  \\
		Sex   & Female &       & 62,652 & (39.3\%) &       & 236,296 & (57.5\%) \\
		&       &       &       &       &       &       &  \\
		Chronic & $\leq$ Stage 2 &       & 88,826 & (55.7\%) &       & 146,825 & (35.7\%) \\
		Kidney & Stage 3a &       & 10,535 & (6.6\%) &       & 15,489 & (3.8\%) \\
		Disease & Stage 3b &       & 2,728  & (1.7\%) &       & 3,127  & (0.8\%) \\
		Stage & Stage 4 &       & 457   & (0.3\%) &       & 551   & (0.1\%) \\
		& \emph{Missing} &       & 56,843 & (35.7\%) &       & 245,205 & (59.6\%) \\
		&       &       &       &       &       &       &  \\
		Ethnicity & White &       & 63,791 & (40.0\%) &       & 153,747 & (37.4\%) \\
		& South Asian &       & 3,072  & (1.9\%) &       & 4,734  & (1.2\%) \\
		& Black &       & 1,065  & (0.7\%) &       & 3,905  & (0.9\%) \\
		& Mixed &       & 237   & (0.1\%) &       & 681   & (0.2\%) \\
		& Other &       & 814   & (0.5\%) &       & 1,623  & (0.4\%) \\
		& \emph{Missing} &       & 90,410 & (56.7\%) &       & 246,507 & (59.9\%) \\
		&       &       &       &       &       &       &  \\
		Comorbidities: &       &       &       &       &       &       &  \\
		Diabetes Mellitus & Yes   &       & 44,727 & (28.1\%) &       & 38,714 & (9.4\%) \\
		Ischaemic Heart Disease & Yes   &       & 42,214 & (26.5\%) &       & 76,013 & (18.5\%) \\
		Arrhythmia & Yes   &       & 17,494 & (11.0\%) &       & 39,094 & (9.5\%) \\
		Cardiac Failure & Yes   &       & 18,647 & (11.7\%) &       & 13,074 & (3.2\%) \\
		Hypertension & Yes   &       & 124,340 & (78.0\%) &       & 240,135 & (58.4\%) \\
		&       &       &       &       &       &       &  \\
		Other anti-hypertensives: &       &       &       &       &       &       &  \\
		Beta-blocker & Yes   &       & 14,666 & (9.2\%) &       & 205,156 & (49.9\%) \\
		Calcium Channel Blocker & Yes   &       & 3,501  & (2.2\%) &       & 91,912 & (22.4\%) \\
		Diuretic & Yes   &       & 21,950 & (13.8\%) &       & 129,582 & (31.5\%) \\
		\hline
	\end{tabular}%
	\captionsetup{margin=20pt,justification=justified,singlelinecheck=false}
	\vspace{0.15in}
	\caption*{Abbreviations: \\ ACEI/ARBs: Angiotensin-converting enzyme inhibitors and angiotensin receptor blockers \\  Diuretic: Thiazide diuretics, Loop diuretics or Potassium sparing diuretics}
\end{table}%

\newpage
\begin{table}[h]
	\centering
	\caption{Estimated effects of ACEI/ARBs on AKI using inverse-probability of treatment weighting (IPTW) to account for confounding.}
	\label{tab:akires}
	\begin{tabular}{cccc}
		\hline
		Confounder &Missing data & Risk difference & Normal-based \\
		adjustment &method &(per 1000 people) &bootstrap 95\% CI \\
		\hline
		Crude &None & 13.30  & (12.52, 14.08) \\
		IPTW &Complete Case Analysis   & 4.60  & (2.76, 6.45) \\
		IPTW &Missingness Pattern Approach  & 5.96  & (5.10, 6.82) \\
		IPTW &Missing Indicator Approach & 5.93  & (5.01, 6.85) \\
		IPTW &Multiple Imputation & 6.17  & (5.27, 7.07)$^{\ast}$ \\
		\hline
	\end{tabular}
	\caption*{\footnotesize{$^{\ast}$ Not bootstrapped; obtained by using Rubin's rules across 10 imputed datasets.}}
\end{table}

\newpage
\begin{table}[htbp]
	\centering
	\caption{Standardised mean differences of confounders, before and after inverse probability of treatment weighting for complete records analysis (CRA), missingness pattern approach (MPA), missing indicator approach (MIndA), and multiple imputation (MI). \newline A standardized difference greater than 10\% indicates imbalance for that variable. \newline ($^{\ast}$ Standardized differences for multiple imputation were averaged over 10 imputed datasets.)}
	\label{tab:akibal}
	\begin{tabular}{llrrrrr}
		\hline
		&      \multicolumn{6}{r}{Percentage standardized differences (absolute values)} \\
		&       & In original & After & After & After & After  \\
		Covariate &  & sample & CRA & MPA & MIndA & MI$^{\ast}$  \\
		\hline
		Age (years)   & 18 to 42 &       &       &       &       &     \\
		& 43 to 53 & 14.64 & 1.33  & 0.49  & 0.36  & 0.26 \\
		& 54 to 62 & 9.42  & 1.64  & 1.51  & 1.45  & 2.26  \\
		& 63 to 71 & 2.48  & 2.17  & 2.11  & 1.98  & 2.93 \\
		& $\geq$ 72 & 4.07  & 2.25  & 3.70   & 3.60   & 4.81 \\
		&       &       &       &       &       &      \\
		Sex  & Female & 36.95 & 1.92  & 4.44  & 4.99  & 4.66  \\
		&       &       &       &       &       &      \\
		Chronic  & $\leq$ Stage 2 &       &       &       &       & \\
		Kidney & Stage 3a & 12.84 & 1.77  & 1.13  & 1.08  & 1.27 \\
		Disease & Stage 3b & 8.62  & 0.07  & 0.35  & 0.38  & 4.56 \\
		& Stage 4 & 3.33  & 0.32  & 0.19  & 0.16  & 1.19  \\
		&       &       &       &       &       &      \\
		Ethnicity & White &       &       &       &       &   \\
		& South Asian & 6.31  & 0.38  & 0.65  & 0.65  & 7.63 \\
		& Black & 3.14  & 3.75  & 3.75  & 4.22  & 8.30 \\
		& Mixed & 1.73  & 0.56  & 0.69  & 0.84  & 4.25   \\
		& Other & 0.43  & 0.42  & <0.01  & 0.01  & 1.12  \\
		&       &       &       &       &       &      \\
		\multicolumn{2}{l}{Diabetes Mellitus} & 49.21 & 0.43  & 2.70   & 2.01  & 3.06  \\
		\multicolumn{2}{l}{Ischaemic Heart Disease}  & 19.25 & 2.52  & 2.83  & 2.28  & 6.00     \\
		\multicolumn{2}{l}{Arrhythmia} & 4.84  & 0.68  & 2.16  & 3.07  & 2.04 \\
		\multicolumn{2}{l}{Cardiac Failure}  & 32.90  & 1.75  & 0.02  & 0.03  & 0.19  \\
		\multicolumn{2}{l}{Hypertension} & 43.08 & 5.74  & 7.85  & 7.88  & 10.93  \\
		\hline						
	\end{tabular}%
\end{table}%

\clearpage

\begin{figure}[h]
	\begin{centering}
		\includegraphics[width=0.3\textwidth]{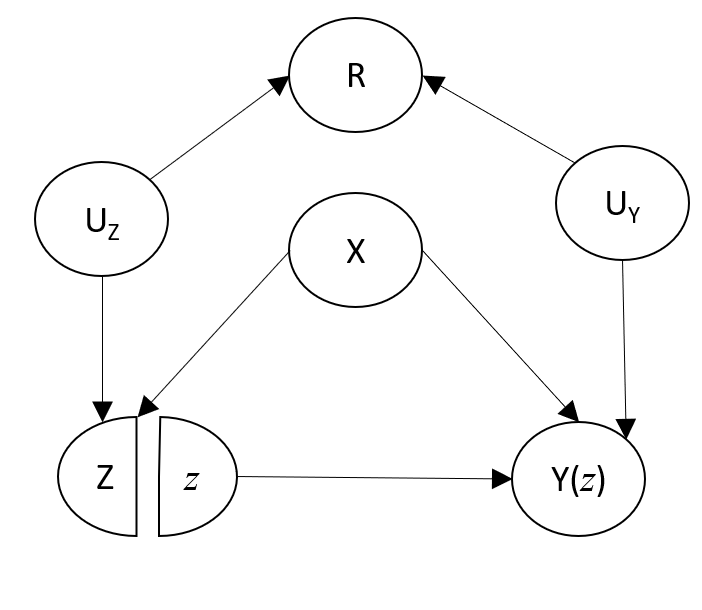}
		\caption{A simple causal diagram showing a scenario in which the mSITA assumption is violated. \newline { $X$: confounder. $Z$: treatment. $Y(z)$: potential outcome resulting from intervening to set $Z$ equal to a particular value $z$. $R$: missingness indicator (=1 if $X$ observed, =0 if $X$ is missing). $U_Z$: unobserved common cause between $R$ and $Z$. $U_Y$: unobserved common cause between $R$ and $Y(z)$.}}\label{fig:dagintro}
	\end{centering}
\end{figure}

\clearpage

\clearpage

\clearpage

\clearpage

\begin{figure}[h]
	\begin{centering}
		\includegraphics[width=1\textwidth]{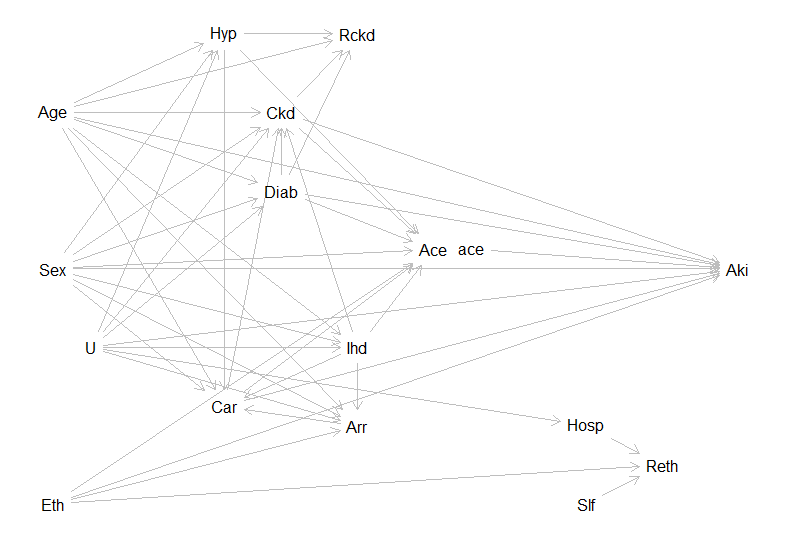}
		\caption{Causal diagram (SWIT) for the motivating example. \\ { Eth: Ethnicity. Ckd: Baseline chronic kidney disease. Hyp: Hypertension. Diab: Diabetes. Arr: Arrhythmia. Car: Cardiac failure. Ihd: Ischaemic heart disease. Ace: Prescription of ACE/ARBs (treatment). ace: intervened-on version of exposure. Aki: Acute kidney injury (outcome). Rckd: Missingness of Ckd. Reth: Missingness of Eth. Hosp: Hospitalisation. Slf: Service-level factors. U: unmeasured factor.}}\label{fig:akiexample} 
	\end{centering}
\end{figure}

\clearpage

\begin{figure}[h]
	\begin{centering}
		\includegraphics[width=0.3\textwidth]{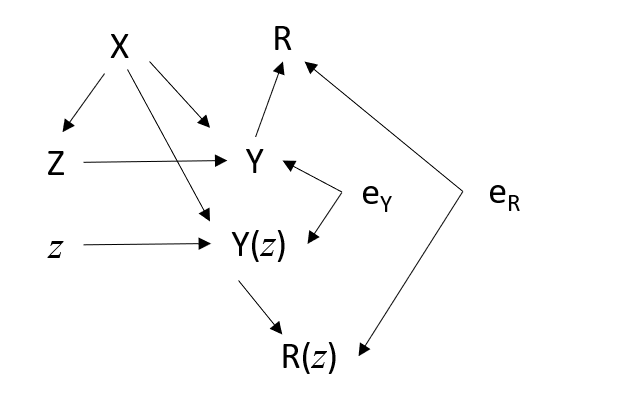}
		\caption{A simple twin network. \\ 
		{ $X$: partially observed confounder. $Z$: observed treatment allocation. $Y$: observed outcome. $Y(z)$: potential outcome resulting from intervening to set treatment to value $z$. $R$: observed missingness indicator (=1 if $X$ observed, =0 if $X$ is missing). $R(z)$: potential missingness indicator (=1 if $X$ observed in counterfactual world, =0 if $X$ is missing in counterfactual world). $e_Y$: unobserved error term between $Y$ and $Y(z)$. $e_R$: unobserved error term between $R$ and $R(z)$.
		}}\label{fig:DAGtn}
	\end{centering}
\end{figure}

\clearpage

\begin{figure}[h]
	\begin{centering}
		\includegraphics[width=1\textwidth]{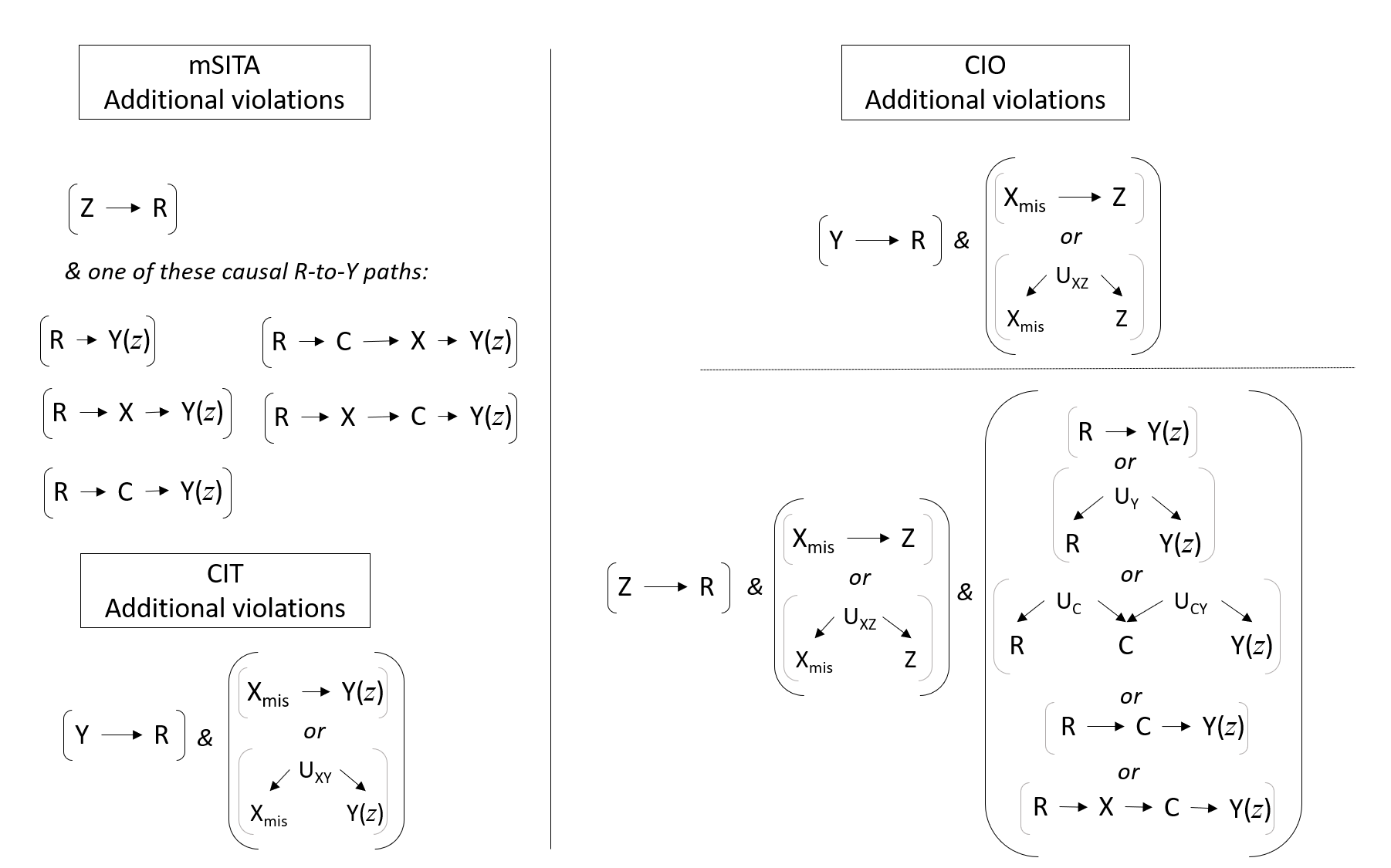}
		\caption{Summary of additional violations of the mSITA, CIT and CIO assumptions. \newline $X$: partially observed confounder. $X_{mis}$: unobserved confounder values. $C$: fully observed confounder. $Z$: treatment. $Y(z)$: potential outcome resulting from intervening to set $Z$ equal to a particular value $z$. $Y$: observed outcome. $R$: missingness indicator (=1 if $X$ observed, =0 if $X$ is missing). $U_{st}$: unobserved common cause between two variables $s$ and $t$. $U_{s}$: unobserved common cause between $R$ and another variable $s$. }\label{fig:assumptions_extra}
	\end{centering}
\end{figure}

\clearpage

\appendix

\section{Validity of the MPA}\label{sec:appx:weak}
In this appendix, we demonstrate that $E\Big[\frac{ZY}{e^{\ast}}\Big]=E[Y(1)]$ under the weaker versions of the assumptions presented in the text. 

First, using the consistency assumption and rearranging, we have that: \begin{align}
E\bigg[\frac{ZY}{e^{\ast}}\bigg] =E\bigg[\frac{ZY(1)}{e^{\ast}}\bigg]  
& =E\bigg[E\bigg[\frac{ZY(1)}{e^{\ast}}|X_{obs},R\bigg]\bigg] \notag\\ 
& =E\bigg[\frac{1}{e^{\ast}}E\big[ZY(1)|X_{obs},R\big]\bigg], \label{eq:befweak}
\end{align}
where $e^{\ast}=E[Z|X_{obs},R]$.

Switching briefly to summation notation:
\begin{align*}
E\big[ZY(1)|X_{obs},R\big] & = \sum\sum zyP(Z|X_{obs},R)P(Y(1)|Z,X_{obs},R) \notag\\
& = \sum\sum zyP(Z|X_{obs},R) \sum P(Y(1),X_{mis}|Z,X_{obs},R) \notag\\
& = \sum\sum\sum zyP(Z|X_{obs},R)P(Y(1)|Z,X_{mis},X_{obs},R)P(X_{mis}|Z,X_{obs},R) \notag
\end{align*}
Using mSITA ($Z \perp Y(z) | X, R$ for $z=0,1$) and CIT ($Z \perp X_{mis} | X_{obs}, R$), we have:
\begin{align*}
E\big[ZY(1)|X_{obs},R\big] & = \sum\sum\sum zyP(Z|X_{obs},R)P(Y(1)|X_{mis},X_{obs},R)P(X_{mis}|X_{obs},R) \notag\\
& = \sum\sum zyP(Z|X_{obs},R) \sum P(Y(1),X_{mis}|X_{obs},R) \notag\\
& = \sum\sum zyP(Z|X_{obs},R)P(Y(1)|X_{obs},R) \notag\\
& = E\big[Z|X_{obs},R\big]E\big[Y(1)|X_{obs},R\big]
\end{align*}
We can also show that $E\big[ZY(1)|X_{obs},R\big]=E\big[Z|X_{obs},R\big]E\big[Y(1)|X_{obs},R\big]$ using mSITA with CIO ($Y(z) \perp X_{mis} | X_{obs}, R$ for $z=0,1$) in a similar manner. Thus, we can rewrite equation \ref{eq:befweak} as follows:
\begin{align*}
E\bigg[\frac{ZY}{e^{\ast}}\bigg] & =E\bigg[\frac{1}{e^{\ast}}E\big[Z|X_{obs},R\big]E\big[Y(1)|X_{obs},R\big]\bigg]. \label{eq:aftweak}
\end{align*}

Since $e^{\ast}=E[Z|X_{obs},R]$:
\begin{align*}
E\bigg[\frac{ZY}{e^{\ast}}\bigg] & =E\bigg[E\big[Y(1)|X_{obs},R\big]\bigg] =E[Y(1)].
\end{align*}

Similarly, we can show that $E[(1-Z)Y/(1-e^{\ast})] = E[Y(0)]. \hfill \qed$

%\clearpage

\section{Twin networks}\label{sec:appx:TN}
When considering scenarios in which treatment, or the outcome, has a causal effect on missingness, by construction, the SWITs now include $R(z)$ instead of $R$. 
This means that the SWIT can no longer be used to test the MPA's assumptions. Instead, we can construct twin networks to check such scenarios, as described by Balke and Pearl (1994) \cite{Bal94} and Shpitser and Pearl (2007) \cite{Shp07}.

Briefly, a twin network can be constructed from a directed acyclic graph, which involves real world variables and relationships, by adding counterparts of variables and relationships in the counterfactual world where treatment has been intervened upon to be set to some realisation of the random variable $Z$. 

For example, Figure \ref{fig:DAGtn} shows a simple twin network of a scenario where the confounder $X$ has a causal effect on both treatment and outcome, treatment has a causal effect on outcome, and outcome has a causal effect on missingness of the confounder. The `real world' is shown by $Z$, $Y$, and $R$. The nodes $z$, $Y(z)$, and $R(z)$ show the counterfactual world - what would occur if we set treatment to value $z$. The observed outcome $Y$ and potential outcome $Y(z)$ are connected by an unobserved error term $e_Y$. Similarly, the observed missingness indicator $R$ and potential missingness indicator $R(z)$ are connected by an unobserved error term $e_R$. Because $X$ has a causal effect on outcome, it also has a causal effect on the potential outcome. It does not, however, affect the intervened-on value of treatment, $z$. 

To assess mSITA in this twin network diagram, we need to assess whether $Z \perp Y(z) | R, X$. Conditioning on $X$ blocks the confounding pathway between $Z$ and $Y(z)$. There is a closed path $Z \rightarrow Y \leftarrow e_Y \rightarrow Y(z)$, blocked because $Y$ is a collider on this path. However, conditioning on $R$ opens this path, because conditioning on a descendant of a collider (i.e. something affected by the collider) has a similar, but weaker, effect as conditioning on the collider itself. Thus the path $Z \rightarrow Y \leftarrow e_Y \rightarrow Y(z)$ is open, after conditioning on $R$, so the mSITA assumption may not be appropriate here.  \footnote{As d-separation for twin networks is not complete \cite{SWIGs}, caution should be used in considering the plausibility of results that suggest two variables are not d-separated. }

Dagitty can be used to assess the assumptions in twin networks, just as for SWITs.

%\clearpage

\section{Additional violations of assumptions} \label{sec:appx:assumptions_extra}

Figure \ref{fig:assumptions_extra} summarises additional violations of the MPA's assumptions.

%\clearpage

\section{Using Dagitty to assess the MPA's assumptions} \label{sec:appx:dagitty}

\subsection{Simple example: R code to use Dagitty to assess the MPA's assumptions}\label{sec:appx:dagitty_simple}
Run in R 3.4.0 \cite{Rpackage}, the R code below reads in our causal diagram for Figure \ref{fig:dagintro} and uses d-separation to assess the mSITA, CIT and CIO assumptions \cite{DAGitty}.

\begin{verbatim}

### R CODE TO USE DAGITTY: SIMPLE EXAMPLE

install.packages("dagitty")
library("dagitty")

############################
#   Load DAG into Dagitty  #
############################

# X       partially observed confounder
# R       observed covariate indicator: =1 if X observed, =0 otherwise.
# Z       treatment allocation (fully observed)
# Yz      potential outcome that would be observed when Z=z   
# U_Y     unobserved common cause of R and Y 
# U_Z     unobserved common cause of R and Z 

g1 <- dagitty( 'dag {
z -> Yz
Z <- X -> Yz
R <- U_Z -> Z
R <- U_Y -> Yz
}')
coordinates( g1 ) <-
list( x=c(Z=1, z=1.2, X=2, Yz=3, R=2, U_Z=1.2, U_Y=2.8),
y=c(Z=3, z=3,   X=2, Yz=3, R=1, U_Z=1.7, U_Y=1.7) )
plot( g1 )

### Assess mSITA assumption:  
###		- Is Z indep of Yz given R, X, and z?
###      (note: we add z to the conditioning set because we are using a SWIG)

# List all paths between Z and Yz
paths( g1, "Z", "Yz", c("R","X","z") )

# Check whether mSITA holds
dseparated( g1, "Z", "Yz", c("R","X", "z") )

# Check whether mSITA holds if U_Y were also measured and 
# included in the confounder set
dseparated( g1, "Z", "Yz", c("R","X", "z", "U_Y") )


#########################################
#   DAG for subgroup with X unmeasured  #
#########################################

### Suppose we believe that X does not effect Z when unmeasured
### R is now written R0 as shorthand for "R=0"

g2 <- dagitty( 'dag {
z -> Yz
X -> Yz
R0 <- U_Z -> Z
R0 <- U_Y -> Yz
}')
coordinates( g2 ) <-
list( x=c(Z=1, z=1.2, X=2, Yz=3, R0=2, U_Z=1.2, U_Y=2.8),
y=c(Z=3, z=3,   X=2, Yz=3, R0=1, U_Z=1.7, U_Y=1.7) )
plot( g2 )

### Assess CIT assumption:   
###		- Is Z indep of X given R=0  (and z)?
###      (note: we add z to the conditioning set because we are using a SWIG)

dseparated( g2, "Z", "X", c("R0", "z") )
paths( g2, "Z", "X", c("R0","z"))

### Assess CIO assumption:   
###		- Is Yz indep of X given R=0  (and z)?
###      (note: we add z to the conditioning set because we are using a SWIG)

dseparated( g2, "Yz", "X", c("R0", "z") )
paths( g2, "Yz", "X", c("R0","z"))
\end{verbatim}

%\clearpage

\subsection{Motivating example: R code to use Dagitty to assess the MPA's assumptions}\label{sec:appx:dagitty_aki}
Figure \ref{fig:akiexample} shows the causal diagram which represents what the investigators believe to represent the underlying causal structure giving rise to the data. The R code below reads in our causal diagram for our motivating example and uses d-separation to assess the mSITA, CIT and CIO assumptions.

%\clearpage

\begin{verbatim}

### R CODE TO USE DAGITTY: MOTIVATING EXAMPLE

#install.packages("dagitty")
library("dagitty")

############################
#   Load DAG into Dagitty  #
############################

# Outcome and treatment:
#   Aki   Acute Kidney Injury (outcome)
#   Ace   ACE/ARB (treatment)
#   ace   Intervened-on ACE/ARB (intervened-on treatment)

# Partially observed confounders and missingness indicators:
#   Eth   Ethnicity (partially observed confounder)
#   Ckd   Baseline CKD (partially observed confounder)
#   Reth  Missingness of ethnicity
#   Rckd  Missingness of baseline CKD

# Determinants of missing data:
#   Slf   Service-level factors determining whether or not ethnicity is measured
#   Hosp  Hospitalisation

# Fully observed confounders:
#   Age   
#   Sex  
#   Hyp   Hypertension
#   Diab  Diabetes
#   Arr   Arrhythmia
#   Car   Cardiac failure
#   Ihd   Ischaemic heart disease

# Unmeasured factors:
#   U     (e.g. frailty)

###   Draw DAG  ###
g1 <- dagitty( 'dag {
Age -> Hyp Age -> Diab Age -> Ckd Age -> Arr Age -> Car Age -> Ihd
Sex -> Hyp Sex -> Diab Sex -> Ckd Sex -> Arr Sex -> Car Sex -> Ihd
Reth <- Eth Reth <- Slf Reth <- Hosp
Rckd <- Hyp Rckd<- Ckd Rckd <- Diab Rckd <- Age
Diab -> Ckd Ihd -> Ckd Car -> Ckd 
Eth -> Arr Ihd -> Arr Arr -> Car Hyp -> Car Ihd -> Car
U -> Ckd U -> Hyp U -> Diab U -> Hosp U -> Ihd U-> Arr
Hyp -> Ace Sex -> Ace Diab -> Ace Eth -> Ace Ckd -> Ace Car -> Ace Ihd -> Ace 
Age -> Aki Eth -> Aki Sex -> Aki Diab -> Aki Ckd -> Aki U -> Aki Car -> Aki
ace -> Aki
}')
coordinates( g1 ) <-
list( x=c(Age=1, Sex=1, Eth=1, Ace=6,    ace=6.5,  Aki=10, 
Arr=5, Car=3.25, Ihd=5,
Reth=9,   Slf=8, Hosp=8, Hyp=3.25, Ckd=4, Diab=4, Rckd=5, U=1.5),
y=c(Age=3, Sex=5, Eth=8, Ace=4.75, ace=4.75, Aki=5, 
Arr=7, Car=6.75, Ihd=6,
Reth=7.5, Slf=8, Hosp=7, Hyp=2,    Ckd=3, Diab=4, Rckd=2, U=6) )
plot( g1 )


#############################
#   Check mSITA assumption  #
#############################

### mSITA assumption:  
###   		Is Z indep of Yz given R, X, and z?
### Here: 	Is Ace indep of Aki given Rckd, Reth, Ckd, Eth, ...
###			...Age, Sex, Hyp, Diab, Arr, Car, Ihd and ace? 
###

# List all paths between Z and Yz
paths( g1, "Ace", "Aki", c("Rckd", "Reth","Ckd", "Eth", 
"Age", "Sex", "Hyp", "Diab",
"Arr", "Car", "Ihd", "ace") )

# Check whether mSITA holds
dseparated( g1, "Ace", "Aki", c("Rckd", "Reth","Ckd", "Eth",
"Age", "Sex", "Hyp", "Diab",
"Arr", "Car", "Ihd", "ace") )


#########################################################
#   DAG for subgroup with CKD and ethnicity unmeasured  #
#########################################################


### Suppose we believe that:
###     Ckd does not affect prescription of ACE when unmeasured
###     Eth does not affect prescription of ACE when unmeasured

###   Draw DAG  (group with neither ethnicity nor ckd measured) ###
g2 <- dagitty( 'dag {
Age -> Hyp Age -> Diab Age -> Ckd Age -> Arr Age -> Car Age -> Ihd
Sex -> Hyp Sex -> Diab Sex -> Ckd Sex -> Arr Sex -> Car Sex -> Ihd
Reth <- Eth Reth <- Slf Reth <- Hosp
Rckd <- Hyp Rckd<- Ckd Rckd <- Diab Rckd <- Age
Diab -> Ckd Ihd -> Ckd Car -> Ckd 
Eth -> Arr Ihd -> Arr Arr -> Car Hyp -> Car Ihd -> Car
U -> Ckd U -> Hyp U -> Diab U -> Hosp U -> Ihd U-> Arr
Hyp -> Ace Sex -> Ace Diab -> Ace Car -> Ace Ihd -> Ace 
Age -> Aki Eth -> Aki Sex -> Aki Diab -> Aki Ckd -> Aki U -> Aki Car -> Aki
ace -> Aki
}')
coordinates( g2 ) <-
list( x=c(Age=1, Sex=1, Eth=1, Ace=6,    ace=6.5,  Aki=10, 
Arr=5, Car=3.25, Ihd=5,
Reth=9,   Slf=8, Hosp=8, Hyp=3.25, Ckd=4, Diab=4, Rckd=5, U=1.5),
y=c(Age=3, Sex=5, Eth=8, Ace=4.75, ace=4.75, Aki=5, 
Arr=7, Car=6.75, Ihd=6,
Reth=7.5, Slf=8, Hosp=7, Hyp=2,    Ckd=3, Diab=4, Rckd=2, U=6) )
plot( g2 )


###############################
#   Check CIT/CIO assumption  #
###############################


### CIT assumption:  
###   Is Z indep of X given R=0  (and z)?
###   Here: is Ace indep of Ckd given Rckd=0 and Reth=0, conditional on:
###		 	Age, Sex, Hyp, Diab  (and ace)?
###   Here: is Ace indep of Eth given Rckd=0 and Reth=0, conditional on: 
###			Age, Sex, Hyp, Diab  (and ace)?

# Check whether CIT holds
dseparated( g2, "Ace", "Ckd", c("Rckd", "Reth", 
"Age", "Sex", "Hyp", "Diab",
"Arr", "Car", "Ihd", "ace") )
dseparated( g2, "Ace", "Eth", c("Rckd", "Reth",
"Age", "Sex", "Hyp", "Diab",
"Arr", "Car", "Ihd", "ace") )

### CIO assumption:   
###   Is Yz indep of X given R=0  (and z)?
###   Here: is Aki indep of Ckd given Rckd=0 and Reth=0, conditional on: 
###		Age, Sex, Hyp, Diab  (and ace)?
###   Here: is Aki indep of Eth given Rckd=0 and Reth=0, conditional on: 
###		Age, Sex, Hyp, Diab  (and ace)?

# Check whether CIO holds
dseparated( g2, "Aki", "Ckd", c("Rckd", "Reth", 
"Age", "Sex", "Hyp", "Diab", 
"Arr", "Car", "Ihd", "ace") )
dseparated( g2, "Aki", "Eth", c("Rckd", "Reth", 
"Age", "Sex", "Hyp", "Diab", 
"Arr", "Car", "Ihd", "ace") )


### Use similar steps to check CIT/CIO in other missingness pattern subgroups
\end{verbatim}

%\clearpage

\section{Balance of confounders in motivating example} \label{sec:appx:covbalance}
In Table \ref{tab:akibal}, we present standardized differences \cite{Wil14introps} calculated to assess the balance of confounders in our motivating example.

\end{document}